%% file: main.tex
\documentclass[lettersize,journal]{IEEEtran}
\usepackage{amsmath,amsfonts}
\usepackage{algorithmic}
\usepackage{algorithm}
\usepackage{array}
\usepackage{textcomp}
\usepackage{stfloats}
\usepackage{url}
\usepackage{verbatim}
\usepackage{graphicx}
\usepackage{amsfonts}
\usepackage{amssymb}
\usepackage{amsmath}
\usepackage{mathtools}
\usepackage{multirow}
\usepackage{tabularx}
\usepackage{arydshln}
\usepackage{hyperref} 
\usepackage{url} 
\usepackage{subcaption}
\usepackage{enumitem}
\usepackage{graphicx}
\usepackage{pdflscape}
\usepackage[justification=centering]{caption}
\usepackage[capitalise]{cleveref}
\usepackage[figuresright]{rotating}
\usepackage{cite}
\usepackage{color}
\usepackage{xcolor}

\begin{document}

\title{Recent Advances in Data-driven Intelligent Control for Wireless Communication: A Comprehensive Survey}

\author{Wei Huo, Huiwen Yang, Nachuan Yang, Zhaohua Yang, Jiuzhou Zhang, Fuhai Nan, Xingzhou Chen, Yifan Mao, Suyang Hu, Pengyu Wang, Xuanyu Zheng, Mingming Zhao, Ling Shi, \IEEEmembership{Fellow,~IEEE}
\thanks{W. Huo, H. Yang, N. Yang, Z. Yang, J. Zhang, F. Nan, X. Chen, Y. Mao, S. Hu, P. Wang, and L. Shi are with the Department of Electronic and Computer Engineering, Hong Kong University of Science and Technology, Clear Water Bay, Hong Kong 
(email: {whuoaa, 
hyangbr, 
nyangaj,
zyangcr,
jzhanghz,
fnan,
xchenfk,
ymaoat,
shuat,
pwangat}@connect.ust.hk,
eesling@ust.hk).}
\thanks{X. Zheng is with Theory Lab, Central Research Institute, 2012 Labs, Huawei Technologies Co. Ltd, Hong Kong Science Park, Hong Kong (email: zhengxuanyu@huawei.com).}
\thanks{M. Zhao is with Noah's Ark Lab, Central Research Institute, 2012 Labs, Huawei Technologies Co. Ltd, Hong Kong Science Park, Hong Kong (email: zhaomingming9@huawei.com).}}

\markboth{Journal of \LaTeX\ Class Files,~Vol.~14, No.~8, August~2021}%
{Shell \MakeLowercase{\textit{et al.}}: A Sample Article Using IEEEtran.cls for IEEE Journals}


\maketitle

\begin{abstract}
The advent of next-generation wireless communication systems heralds an era characterized by high data rates, low latency, massive connectivity, and superior energy efficiency. These systems necessitate innovative and adaptive strategies for resource allocation and device behavior control in wireless networks. Traditional optimization-based methods have been found inadequate in meeting the complex demands of these emerging systems. As the volume of data continues to escalate, the integration of data-driven methods has become indispensable for enabling adaptive and intelligent control mechanisms in future wireless communication systems.
This comprehensive survey explores recent advancements in data-driven methodologies applied to wireless communication networks. It focuses on developments over the past five years and their application to various control objectives within wireless cyber-physical systems.
It encompasses critical areas such as link adaptation, user scheduling, spectrum allocation, beam management, power control, and the co-design of communication and control systems. 
We provide an in-depth exploration of the technical underpinnings that support these data-driven approaches, including the algorithms, models, and frameworks developed to enhance network performance and efficiency.
We also examine the challenges that current data-driven algorithms face, particularly in the context of the dynamic and heterogeneous nature of next-generation wireless networks. The paper provides a critical analysis of these challenges and offers insights into potential solutions and future research directions. This includes discussing the adaptability, integration with 6G, and security of data-driven methods in the face of increasing network complexity and data volume.
%
\end{abstract}

\begin{IEEEkeywords}
Data-driven methods, artificial intelligence, next-generation wireless communication, cyber-physical systems
\end{IEEEkeywords}

\input{Introduction.tex}

\begin{table*}[t] 
	\caption{List of acronyms}
 \centering
	\label{tab: acronyms}
\scriptsize
\begin{tabular}[c c c c]{|m{0.1\textwidth}<{\centering}|m{0.3\textwidth}<{\centering}|m{0.1\textwidth}<{\centering} |m{0.3\textwidth}<{\centering}|}
\hline
Acronym	&  Full meaning & Acronym  & Full meaning \\
 \hline
 AI & Artificial intelligent & 4G & Fourth-generation \\
 \hline
 5G & Fifth-generation & 6G & Sixth-generation \\
 \hline
  ML & Machine learning & DL & Deep learning \\
 \hline
 RL & Reinforcement learning & DRL & Deep reinforcement learning  \\
 \hline
 NN & Neural network & DDPG & Deep deterministic policy gradient \\
 \hline
 KM & K-means & TS & Thompson sampling \\
 \hline
 UCB & Upper confidence bound & SARSA & State-action-reward-state-action \\
 \hline
 DQN & Deep Q-network & OLLA & Outer loop link adaptation \\
 \hline
 ILLA & Inner loop link adaptation & BLER & Block error rate \\
 \hline
 HARQ & Hybrid automatic repeat request & THz & Tera Hertz \\
 \hline
 MINLP & Mixed integer nonlinear programming & MILP & Mixed integer linear programming \\
 \hline
 SU & Secondary user & PU & Primary user \\
 \hline
LP & Linear programming & UDN & Ultra-dense network \\
\hline
 AMC & Adaptive modulation and coding & OFDM & Orthogonal frequency division multiplexing \\
\hline
 MCS & Modulation and coding scheme & SNR & Signal-to-noise ratio \\
\hline
BS & Base station & UE & User equipment \\
\hline
mmWave & Millimeter wave & MIMO & Multiple-input multiple-output \\
\hline
URLLC & Ultra-reliable low latency communications & eMBB & Enhanced mobile broadband \\
\hline
UAV & Unmanned aerial vehicle & ADI & Angular domain information \\
\hline
AoA & Angle-of-arrival & AoD & Angle-of-departure \\
\hline
WNCS & Wireless networked control systems & CPS & Cyber-physical system\\
\hline
IoT & Internet of things & MDP &  Markov decision process \\
\hline
ETC & Event-triggered control & WMN & Wireless mesh network\\
\hline
DoS & Denial-of-service & DDQN & Double deep Q-network \\
\hline
AoI & Age-of-information &  BLER & Block error rate \\
\hline   
SINR & Signal-interference-to-noise ratio & CCI & Co-channel interference \\
\hline
QoS & Quality of service & CSI & Channel state information \\
\hline
DNN & Deep neural network & KNN & K-nearest neighbors\\
\hline
SVM & Support vector machine & GPML & Gaussian process machine learning \\
\hline
SGD & Stochastic gradient descent & ADMM & Alternating direction method of multipliers\\
\hline
CNN & Convolutional neural network & RNN & Recurrent neural network \\
\hline
LSTM & Long short-term memory & GNN & Graph neural network \\
\hline
ResNet & Residual neural network & SCA & Successive convex approximation \\
\hline
C-RAN & Cloud radio access network & WMMSE & Weighted minimum mean square error \\
\hline
AC & Actor-critic & ReLU & Rectified linear unit \\
\hline
MBS & Macro base station & SBS & Small-cell base station \\
\hline
DSRL & Deep SARSA reinforcement learning & HAPS & High-altitude platform station \\
\hline
TBS & Terrestrial base station &  DQL & Deep Q-learning\\
\hline
FL & Federated learning  &  MAB & Multi-armed bandit  \\
\hline
\end{tabular}
\end{table*}

\input{pre.tex}

\input{link_adaptation.tex}

\input{user_schedule.tex}

\input{spectrum_allocation.tex}

\input{beam_management.tex}

\input{power_control}

\input{codesign.tex}




\section{Challenges and Future Research Directions}\label{sec11}
\subsection{Dynamic Adaptability}
In the rapidly evolving communication landscape, the ability of data-driven methods to adapt dynamically to changing conditions is a crucial yet challenging requirement. Maintaining relevance and effectiveness in the face of constantly shifting user needs, preferences, and behavioral patterns requires communication strategies to be highly responsive.

While there has been significant progress in leveraging historical data for analysis and decision-making, these methods can still lead to insights that are no longer reflective of the current reality, with outdated or irrelevant data resulting in suboptimal or erroneous decisions. The next frontier in ML for communication strategies involves developing advanced techniques that can anticipate and react to changes proactively. Autonomous learning and iterative optimization capabilities must evolve to ensure models remain robust and accurate amidst continuous flux.

Real-time adaptation, decision-making, and optimization tasks are increasingly computationally intensive, presenting ongoing challenges to the scalability of data-driven communication systems. Future research should delve deeper into leveraging RL for more nuanced and adaptive decision-making. Additionally, the use of distributed computing must be further optimized to enhance the scalability and responsiveness of these systems. DL techniques also need to be refined to improve the processing of large data volumes and the identification of intricate patterns and trends that inform communication strategies.

Moreover, while self-supervised learning has shown promise in reducing reliance on labeled data and enhancing model adaptability to new environments, it requires further exploration to maximize its potential for continuous learning. Similarly, online learning methodologies must be advanced to enable real-time model updates with incoming data, maintaining relevance and accuracy in dynamically changing environments.

Lastly, despite the progress in real-time data processing technologies such as stream processing and event-driven architectures, there remains a need for more sophisticated methods to incorporate the latest data seamlessly into communication strategies. This will enable more responsive and up-to-date decision-making.

By addressing these deeper challenges and pursuing these advanced research directions, the field of data-driven communication can unlock the full potential of dynamic adaptability, ensuring that communication strategies remain relevant, effective, and responsive in the face of ever-changing user behaviors and environmental conditions.

\subsection{Integration with 6G and Beyond}
As the world looks forward to the rollout of 6G around 2030, the research community is delving into integrating data-driven techniques to address the formidable challenges posed by future wireless communication systems. 6G aspires to deliver unprecedented data rates, ultra-low latency, and ubiquitous connectivity, facilitating next-generation applications such as immersive augmented reality (AR), virtual reality (VR), holographic communications, and the Internet of Everything (IoE). To meet these ambitious goals, key research areas include managing ultra-dense networks (UDNs), optimizing terahertz (THz) communications, and enhancing integrated sensing-communication systems.
Ultra-dense networks are characterized by deploying a vast number of small cells within a limited geographical area to satisfy the high capacity and coverage demands of 6G. This dense deployment introduces significant challenges, including complex interference management, efficient resource allocation, and seamless mobility management. Traditional approaches may fall short in such a dynamic and dense environment. However, ML and AI offer promising solutions. For instance, AI can optimize handover decisions in real time, reducing latency and improving user experience. Predictive analytics can forecast traffic patterns, allowing for more efficient resource allocation and load balancing. RL can dynamically adjust power levels and beamforming strategies to mitigate interference, ensuring robust and reliable communication. By leveraging the massive amounts of data generated by UDNs, these data-driven techniques can significantly enhance network performance and reliability.

THz communications, which operate in the frequency range of 0.1-10 THz, are expected to be a cornerstone of 6G, providing ultra-high data rates and bandwidth. However, THz signals face significant challenges, including high path loss, atmospheric absorption, and limited penetration capabilities. Data-driven approaches can help overcome these obstacles. For example, advanced ML algorithms can develop more accurate channel models by analyzing large datasets of THz channel measurements, which is crucial for effective communication at these frequencies. AI can also optimize beamforming and beam-steering techniques to maintain strong and stable links in dynamic environments. Furthermore, data-driven algorithms can enhance error correction codes and modulation schemes, improving the reliability and efficiency of THz communications. These techniques can adapt to the unique propagation characteristics of THz signals, ensuring that the high potential of THz communications is fully realized in 6G networks.

The integration of sensing and communication functionalities is another significant aspect of 6G. This convergence envisions devices that not only communicate but also sense their environment, supporting applications like smart cities, autonomous driving, and environmental monitoring. This integration requires sophisticated data processing and interpretation capabilities. ML and AI can significantly enhance integrated sensing-communication systems. For instance, ML algorithms can process sensor data to detect patterns and anomalies, enabling predictive maintenance and real-time decision-making. AI can facilitate the fusion of data from multiple sensors, improving the accuracy and reliability of environmental monitoring. Additionally, data-driven techniques can optimize the allocation of sensing and communication resources, ensuring efficient use of the available spectrum and enhancing the overall performance of 6G networks. This synergy between sensing and communication can lead to more intelligent and responsive systems, supporting a wide range of advanced applications.

\subsection{Security Challenges and Future Directions}
While data-driven AI approaches show promising potential for optimizing wireless system parameters and enabling intelligent sensing, communication and control, ensuring security is a critical challenge that needs to be addressed.

A key vulnerability of data-driven AI models is their susceptibility to adversarial attacks, where carefully crafted input perturbations can cause mis-classification or degrade model performance. Some attacks in wireless scenarios are: jamming by injecting corrupted sensor data, spoofing physdoical layer parameters to disrupt link adaptation, manipulating spectrum sensing inputs to cause under-utilization, and misleading user scheduling and beamforming decisions.

Another threat arises from data poisoning attacks, where malicious agents inject crafted samples to model, such as poisoning channel quality indicator data, spoofing positioning data in localization,and manipulating system state observations.

Preserving user privacy and preventing leakage of sensitive information from wireless data is also crucial, since potential disclosure could occur in positioning data for location tracking.

To develop secure and trustworthy data-driven wireless systems, an interdisciplinary research effort spanning communications, AI and security is required. Some future directions include wireless-aware adversarial AI defenses, and privacy-preserving AI models for wireless data.

\section{Conclusion}\label{sec12}
Through our comprehensive review, we emphasize the significance of data-driven methods in in tackling the intricate and ever-evolving nature of wireless communication networks. 
These methods hold tremendous potential in enhancing system performance, optimizing resource allocation, improving spectrum efficiency, and enabling adaptive and intelligent decision-making.
However, further efforts are required to strengthen the security and trustworthy of data-driven wireless systems, enhance the adaptability and generalization of dynamic data-driven algorithms, and seamlessly integrate advanced data-driven algorithms into practical 6G solutions, effectively addressing the challenges faced by the next-generation wireless communication networks.
By gaining a deep understanding of the advancements and challenges in this field, researchers and practitioners can obtain valuable insights into the potential applications and future directions of data-driven intelligent control in wireless communication systems. This knowledge will pave the way for transformative advancements in the field and drive the evolution of wireless communication networks towards greater efficiency and effectiveness.

\bibliographystyle{IEEEtran}
\bibliography{sn-bibliography.bib}

\newpage

 





\end{document}

%% file: Introduction.tex
\section{Introduction}\label{sec1}

Wireless communication systems are the backbone of modern society, enabling applications ranging from personal communication\cite{yu2011resource} to industrial automation \cite{liu2018d2d}. 
Next-generation communication systems are expected to autonomously determine the optimal system configurations by utilizing the intricate and diverse characteristics of both users and environments. 
However, the increasing complexity and demand for higher performance, reliability, and efficiency present challenges that traditional approaches struggle to address \cite{chowdhury20206g}. These conventional techniques often rely on prior knowledge and empirical models, making them inadequate for the rapid changes and complexity inherent in modern wireless environments, such as mobile users and dynamic interference patterns. 
Additionally, traditional optimization-based methods are time-consuming, especially for extremely large systems, failing to provide the necessary scalability and real-time decision-making capabilities required to maintain optimal performance.

Data-driven approaches, encompassing machine learning (ML), deep learning (DL), reinforcement learning (RL), online learning, and other statistical methods, offer significant advantages in overcoming the limitations of traditional methods \cite{ye2020deep}. One of the key strengths of data-driven models is their ability to continuously learn from real-time data, enabling them to adapt dynamically to changing network conditions. This adaptive capability ensures more efficient resource management, such as better spectrum utilization and energy optimization.
In addition to adaptive capabilities, scalability is another critical advantage of data-driven methods. Modern wireless networks, such as those involving massive multiple-input multiple-output (MIMO) or heterogeneous networks \cite{ghazanfari2021model}, present a level of complexity that exceeds the capabilities of traditional optimization techniques. Data-driven approaches can manage this complexity by learning from data patterns that traditional methods cannot handle, enabling automated decision-making processes. This automation reduces the need for manual intervention, increasing operational efficiency and allowing for the development of large-scale applications such as smart cities and the Internet of Things (IoT). 
By improving accuracy and reliability, data-driven approaches pave the way for innovations in these applications.
In summary, while traditional approaches in wireless communication systems are limited by their inability to adapt to dynamic environments and scale with increasing complexity, data-driven approaches offer substantial advantages. 
By leveraging the power of data, these approaches can meet the growing demands and challenges of modern wireless communication networks, ensuring better connectivity, efficiency, and user experience.

\begin{figure*}[t]
    \centering
    \includegraphics[width=0.8\linewidth]{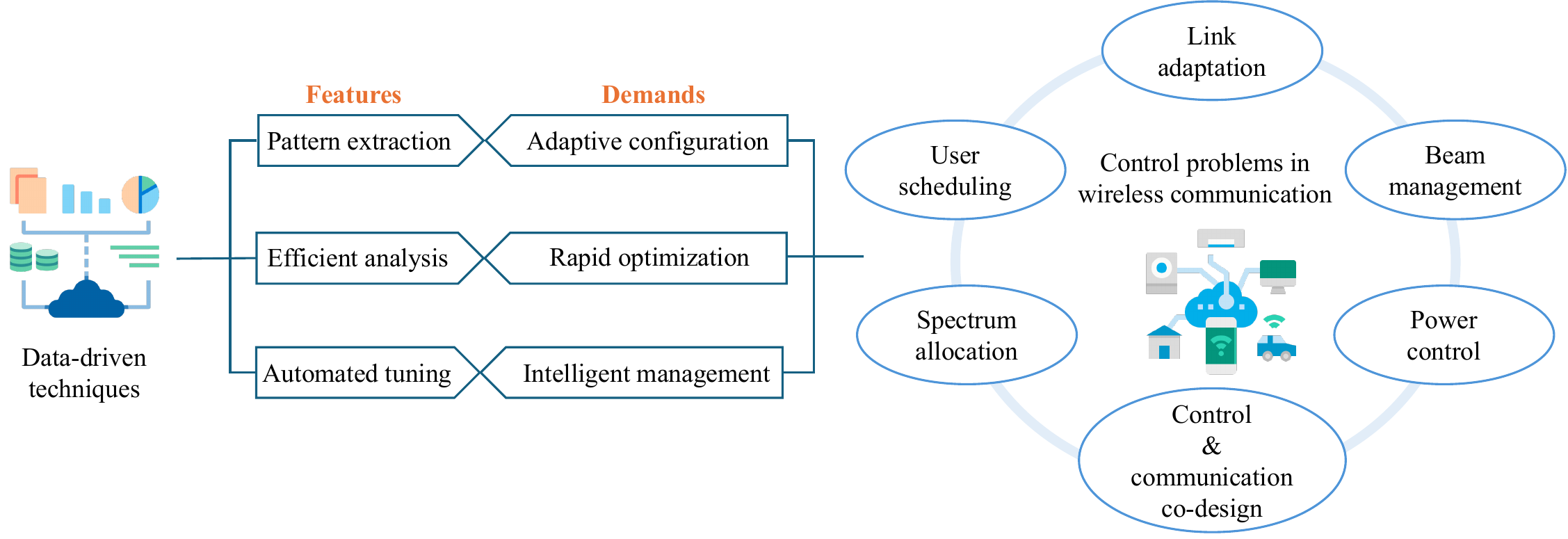}
    \caption{Advantages of data-driven techniques for intelligent control in wireless communication.}
    \label{fig: scope}
\end{figure*}

Given the rapid advancements in both data-driven approaches and communication technologies, this survey aims to provide a comprehensive overview of integrating data-driven techniques into communication systems to address critical issues, enhance performance, and enable new capabilities.
Fig.~\ref{fig: scope} illustrates the scope of this survey. We explore how features inherent in current data-driven techniques, such as pattern extraction, efficient analysis, and automated tuning, facilitate adaptive configuration, rapid optimization, and intelligent management across various control objectives in wireless communication.
Specifically, we review recent advancements in applying data-driven methods to link adaptation, which dynamically adjusts communication link parameters to optimize performance under varying channel conditions. 
We also investigate their role in efficiently allocating network resources to users to maximize throughput and minimize latency. Additionally, we explore how these techniques improve spectrum efficiency and mitigate interference within communication networks. Moreover, we analyze the application of data-driven algorithms in optimizing beamforming and beam management techniques to enhance signal quality, reduce interference, and improve the efficiency of millimeter-wave communications. Furthermore, we discuss the use of data-driven approaches for dynamic power control to optimize energy efficiency while maintaining communication quality. Finally, we examine how big data facilitates the co-design process, enhancing the overall performance and efficiency of networked control systems.

The existing literature on artificial intelligence (AI) and data-driven methodologies for wireless communications is vast, yet often narrowly focused. For instance, Zuo et al.~\cite{zuo2023survey} explored the convergence of blockchain and AI in sixth-generation (6G) communication, yet overlooked critical wireless challenges such as link adaptation and user scheduling. Similarly, Xu et al.~\cite{xu2023state} examined AI-driven backscatter communication, but their analysis was confined to technical integration with fifth-generation (5G)/6G systems, without delving into broader wireless issues.
In contrast to Chen et al.~\cite{chen2019artificial} and Mao et al.~\cite{mao2023deep}, who emphasized the dominance of neural networks, our survey extends the discussion to include the practicality of lightweight ML and online learning in wireless communications, highlighting their potential to streamline processes and enhance efficiency.
While a survey~\cite{hu2021distributed} offered a broad overview of distributed ML in wireless contexts, it bypassed the significance of control aspects in IoT applications, such as the co-design of control and communication. The majority of surveys~\cite{elsayed2019ai, debbabi2022overview, yang2020artificial, ye2024artificial} have centered on information layer resource management, neglecting the synergies between communication and control in cyber-physical systems (CPSs).
Our contribution lies in reviewing the co-design of data-driven control and communication, emphasizing the integration of physical industrial processes with information transmission, which aligns closely with IoT operations. Unlike Mekrache et al.~\cite{mekrache2022deep}, who concentrated on 6G vehicular applications using RL, our survey encompasses a wider array of data-driven approaches, reflecting a more comprehensive view of the field's potential and challenges.
The contributions of this survey are summarized as follows:

\begin{itemize}
    \item This survey comprehensively explores the integration of a variety of data-driven methods, such as ML, DL, online learning, RL, etc., into wireless communication systems. We illustrate how these methods can be utilized to significantly enhance system performance and operational efficiency.
    \item  In wireless communications, our survey explores how data-driven techniques enhance network operations and user experiences in link adaptation, user scheduling, spectrum allocation, beam management, power control, and co-designing communication and control systems. 
    Existing surveys focus solely on information layer resource management, overlooking the synergies in CPSs. Our review integrates physical industrial processes with information transmission, crucial for practical IoT operations.
    \item Recognizing the current limitations and future potential of data-driven methods in wireless communications, our survey identifies key challenges related to adaptability, model complexity, scalability, and security. We propose future research directions aimed at overcoming these obstacles, with the goal of further enhancing the integration of data-driven techniques with communication technologies to achieve more efficient and reliable networks.
\end{itemize}

The remainder of this article is organized as follows. In Section~\ref{sec2}, we present some preliminaries on ML, DL, RL, online learning, and statistical methods. Section~\ref{sec3} - Section~\ref{sec8} is the main body of this survey. Section~\ref{sec3} introduces the topic of link adaptation, discussing its significance and various approaches. Section~\ref{sec4} examines the topic of user scheduling, which is vital for optimizing network performance and user experience. 
Section~\ref{sec5} explores the challenges and solutions related to spectrum allocation, a critical aspect of network management. 
Section~\ref{sec6} delves into beam management, an essential component of modern wireless communication systems. 
Section~\ref{sec7} focuses on power control, detailing its importance in maintaining efficient network operation. 
Finally, Section~\ref{sec8} addresses the integration of communication and control co-design, focusing on its impact on network efficiency and reliability. 
In Section~\ref{sec11}, we further detail the existing challenges and future directions across all the topics above. Finally, we draw a conclusion of the whole survey in Section~\ref{sec12}.
For easy referencing, Table~\ref{tab: acronyms} summarizes the various abbreviations used in this paper.

%% file: pre.tex
\section{Data-driven Frameworks}\label{sec2}
\subsection{Machine Learning}

ML is an important field in AI, which focuses on using statistical methods to explore the relationships between data, and enabling computers to complete tasks when new data emerges. More specifically, ML methods are divided into unsupervised learning and supervised learning, based on whether the models are trained through the labeled dataset. 

K-means (KM) \cite{macqueen1967some} clustering is a popular unsupervised ML algorithm that tends to identify clusters based on the similarity of the provided data under a defined metric or transformation. Each data point is assigned to its nearest centroid, and the \(k\) centroids are subsequently updated by calculating the mean of each cluster. This process iteratively converges to a local optimum, which does not require pre-training. However, the quality of the results relies on the initial centroids and whether value of \(k\) is appropriate. 

Different from KM, the k-nearest neighbors (KNN) \cite{fix1985discriminatory}, support vector machine (SVM) \cite{cortes1995support} and Gaussian processes machine learning (GPML) \cite{rasmussen2003gaussian} are all supervised learning methods. KNN algorithm is an effective non-parametric method used for classification or regression tasks. This algorithm operates by first transforming the given training data into a multidimensional feature space, where each sample is associated with a label. In KNN classification, the label of a test query will be determined by a plurality vote among its \(k\) nearest neighbor samples. On the other hand, in KNN regression, the result for a query data point is computed as the mean of the values from its \(k\) nearest neighbor samples. SVM is widely used for classification tasks. In linear classification, SVM identifies support vectors from the given training data to determine a hyperplane as a decision boundary that effectively separates the data space with distinct labels. For non-linear classification, SVM employs a technique called kernel method, which can transform the given data into higher-dimensional space. By well designing the kernel, a linear separator can be found among the transformed space. GPML is primarily used for regression and probabilistic modeling tasks. Specifically, a Gaussian process is constructed with a mean function and a covariance function (or kernel), which approximates the observed data with uncertainty. As GPML follows the Bayesian framework, both the prior knowledge of the system and new data can be utilized for enhancing model performance.

\subsection{Deep Learning}

\begin{figure}[h]
    \centering
    \captionsetup{justification=centering}
    \includegraphics[scale=0.4]{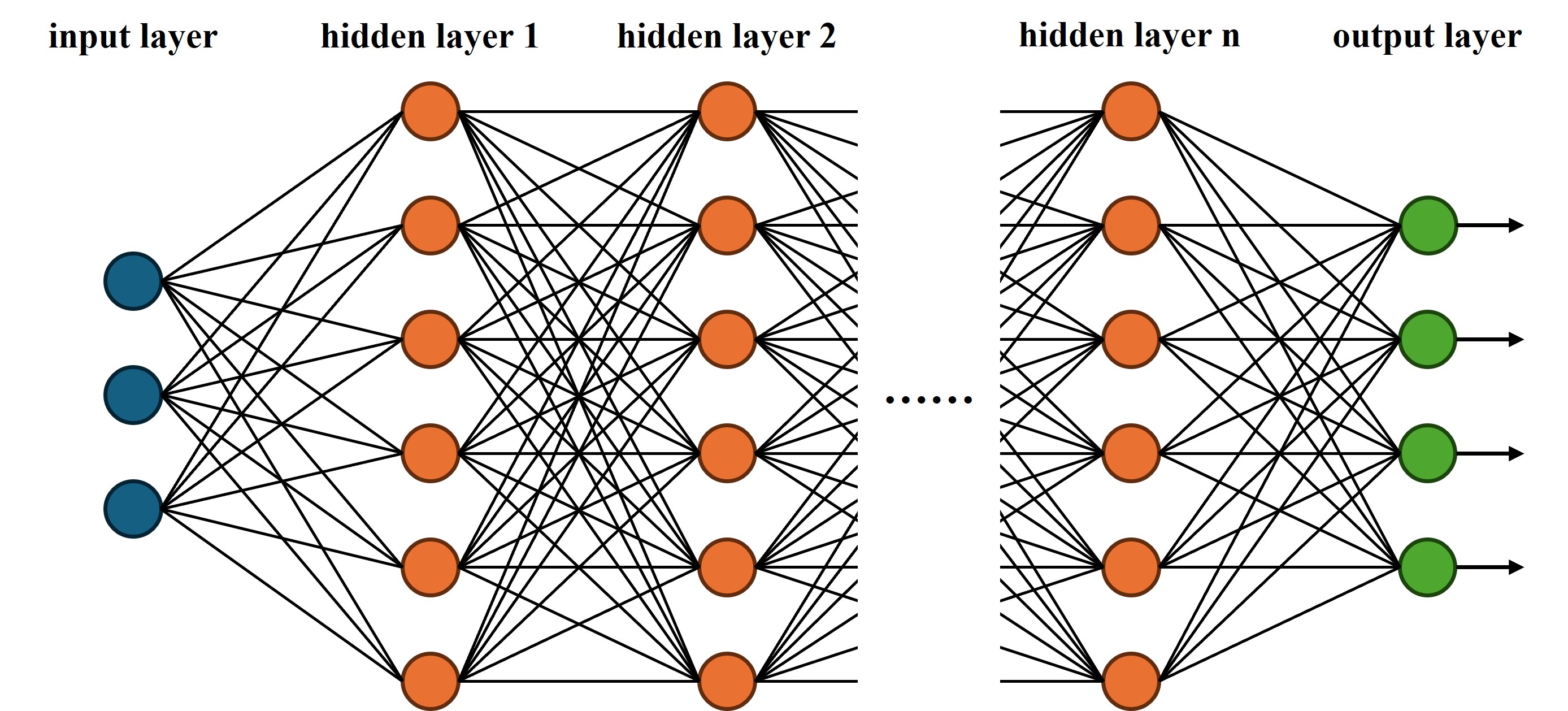}
    \caption{DNN architecture.}
    \label{DL}
\end{figure}

Inspired by the structure and function of human brain, DL is particularly powerful for handling large, complex dataset in the comparison to traditional ML methods \cite{lecun2015deep}. Deep neural networks (DNNs) are the foundation of DL, which explore the connections between input and output data with multiple computational layers. As shown in Fig.~\ref{DL}, a DNN consists of an input layer, hidden layers, and an output layer, where the number of hidden layers can be large for accuracy improvement. The term ''deep'' typically refers to numerous hidden layers in the network. In a DNN, the computation between each layer involves a series of linear transformations followed by activation functions such as \textit{rectified linear unit} (\(\textit{ReLU}\)), \(\textit{sigmoid}\), \(\textit{tanh}\), etc. These activation functions introduce non-linearity into the networks, enabling DNNs to model complex relationships between input and output. The training of DNNs adjusts the parameters within hidden layers to minimize the error between the predicted output and the actual data, which is realized by backpropagation combined with optimization algorithms such as stochastic gradient descent (SGD) and alternating direction method of multipliers (ADMM). Based on CNN, more complex and powerful network structures have been created to enhance DL performance, such as convolutional neural networks (CNN) \cite{krizhevsky2012imagenet}, recurrent neural networks (RNNs) \cite{rumelhart1986learning}, graph neural networks (GNNs) \cite{scarselli2008graph}, residual neural network (ResNet) \cite{he2016deep}, etc.

\begin{figure}[h]
    \centering
    \captionsetup{justification=centering}
    \includegraphics[scale=0.38]{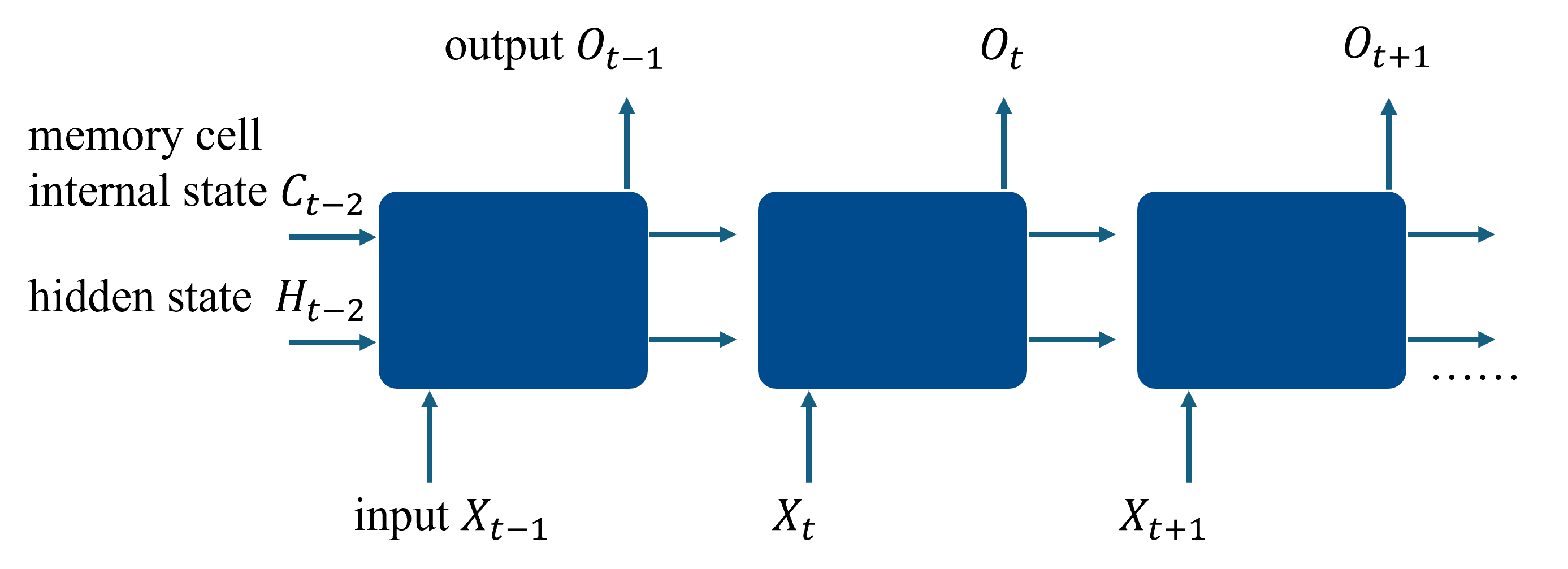}
    \caption{LSTM network architecture.}
    \label{LSTM}
\end{figure}

CNN is widely used for processing grid data like images. By incorporating convolutional kernels, the internal layers of CNNs can effectively learn to extract relevant data features, thereby enabling CNNs to outperform traditional DNNs in various tasks. RNNs belong to bi-directional neural network (NN) architectures, which are capable of handling sequential data. More specifically, unlike traditional feedforward NN, RNNs have cyclic connections, allowing information to persist and be used across different steps in the sequence. Among RNNs, the long short-term memory (LSTM) network is one of the most popular modern architectures, which is shown in Fig.~\ref{LSTM}. In addition to the hidden state being transmitted, the LSTM introduces a memory cell \(C_t\) to store historical information along each time step \(t\), making the network effective for tasks requiring long-term memory.

\begin{figure}[h]
    \centering
    \captionsetup{justification=centering}
    \includegraphics[scale=0.45]{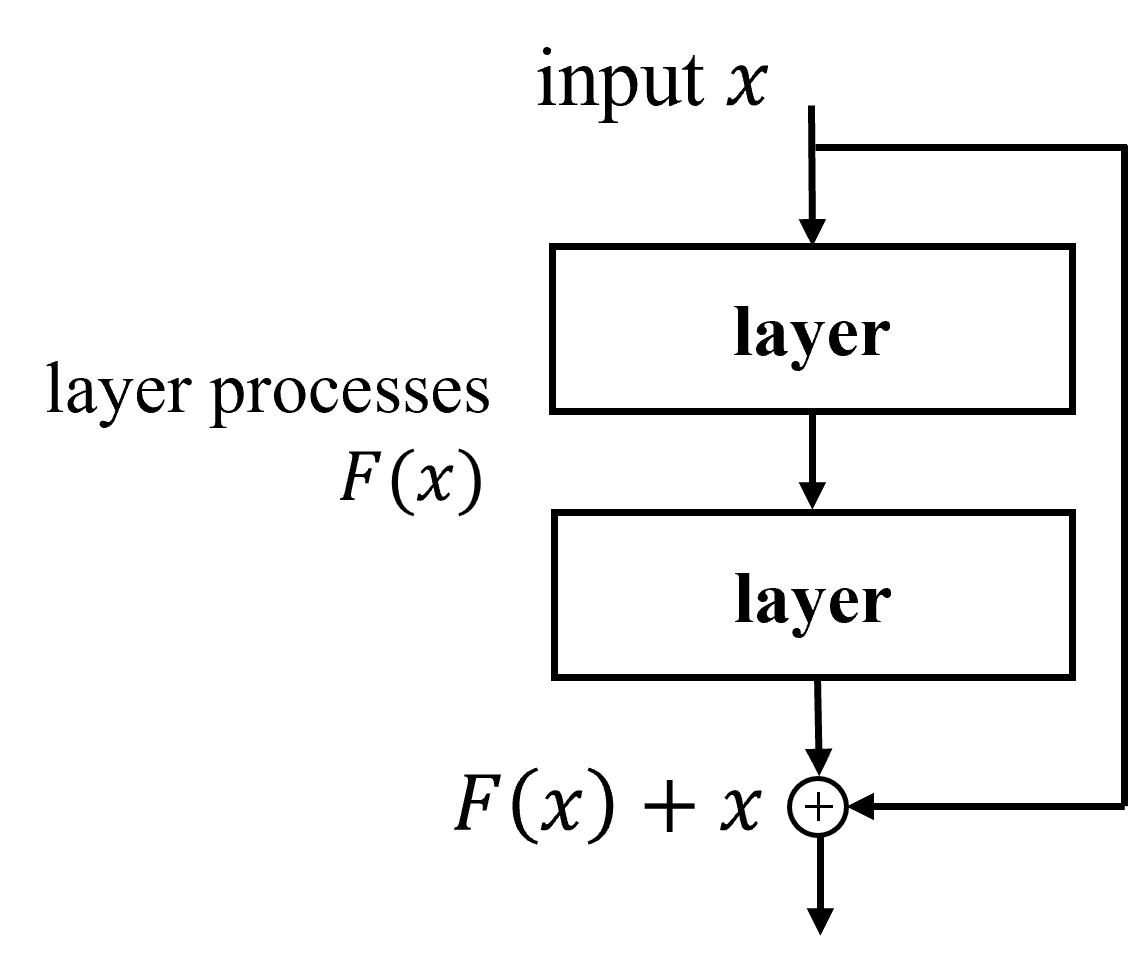}
    \caption{The residual block in ResNet.}
    \label{RES}
\end{figure}

GNNs are designed for modeling graph-structured data. Different from traditional NNs that process data with a fixed structure, GNNs can handle data represented as graphs, which consist of nodes and edges. As communication networks can be naturally described as graphs, the GNNs are highly suitable for modeling communication systems. What's more, ResNet is a famous DNN architecture introduced by He~et~al.~\cite{he2016deep}. The core idea of ResNet is to learn residual functions with reference to the layer inputs. As shown in Fig.~\ref{RES}, a residual block is the basic building unit of ResNet, which consists of several layers with a shortcut connection that skips one or more layers. By introducing residual connections, the networks can mitigate the vanishing gradient problem and be easily scaled, which leads to improved performance and flexibility.

ML and DL excel at capturing the inherent complex dynamics of data. Consequently, these methods effectively establish intricate mappings between varying wireless network conditions and optimal transmission decisions. For instance, they can be employed to learn the look-up table that maps channel states to the best modulation and coding scheme (MCS) index for link adaptation. 
ML and DL techniques can be utilized for control and communication co-design to predict the optimal control strategies and communication parameters based on real-time network conditions, thereby enhancing system efficiency and performance. DL-based methods can learn complex, data-driven power control models that outperform traditional optimization-based techniques, especially in nonlinear, dynamic environments. DL also enables rapid deployment across diverse network topologies through transfer learning, improving scalability. 

\subsection{Online Learning}
Compared to other learning methods, online learning is characterized by its simplicity, adaptability, and ability to make sequential decisions under uncertainty. Online learning is especially useful in scenarios where the number of possible actions is large and the learner must quickly identify the best options without exhaustive exploration. 

Thompson sampling (TS) \cite{thompson1933likelihood} and upper confidence bound (UCB) \cite{cox1992statistical} are two well-known online learning paradigms. TS is an algorithm for sequential decision-making, particularly effective in multi-armed bandit (MAB) problems. It harmonizes exploration and exploitation through probabilistic sampling. TS operates by maintaining a posterior distribution that encapsulates the algorithm's beliefs about the unknown parameters' values. At each step, it samples from this distribution to estimate expected rewards for each action, then chooses the action with the highest estimated reward. After an action's outcome is observed, TS refines its beliefs by updating the posterior with Bayes' rule. This iterative process sharpens the decision-making as more data is accrued. 
UCB is another algorithm for the MAB problem that balances exploration and exploitation by calculating an upper confidence limit for the expected reward of each arm. The UCB value reflects both the estimated reward and the uncertainty of that estimate. At each step, the arm with the highest UCB is chosen, encouraging the selection of arms with greater potential or unknown potential. As data accumulates, the UCB algorithm updates the confidence intervals, which shrink with more observations, reducing uncertainty. This process leads to a progressive preference for arms with higher expected rewards, as the algorithm converges to more informed decisions over time.

\subsection{Reinforcement Learning}
RL differs from other learning paradigms in learning the optimal state-action policy. This is done by constantly interacting with the environment to accumulate the greatest possible reward over time. Unlike supervised learning, which requires labeled examples, RL agents learn by trial and error, balancing the exploration of new actions against the exploitation of known successful ones. This approach allows RL to excel in sequential decision-making problems where delayed feedback is common and the environment may be non-stationary. Moreover, RL does not rely on pre-defined features, as agents learn to extract relevant information from the environment itself, making it adaptable to a wide range of applications from robotics to game playing. While RL can be sample-inefficient and face challenges in scalability, its ability to learn from the consequences of actions and adapt to changing environments sets it apart as a powerful tool for complex, goal-oriented tasks.

\begin{figure} [h]
    \centering
    \captionsetup{justification=centering}
    \includegraphics[scale=0.55]{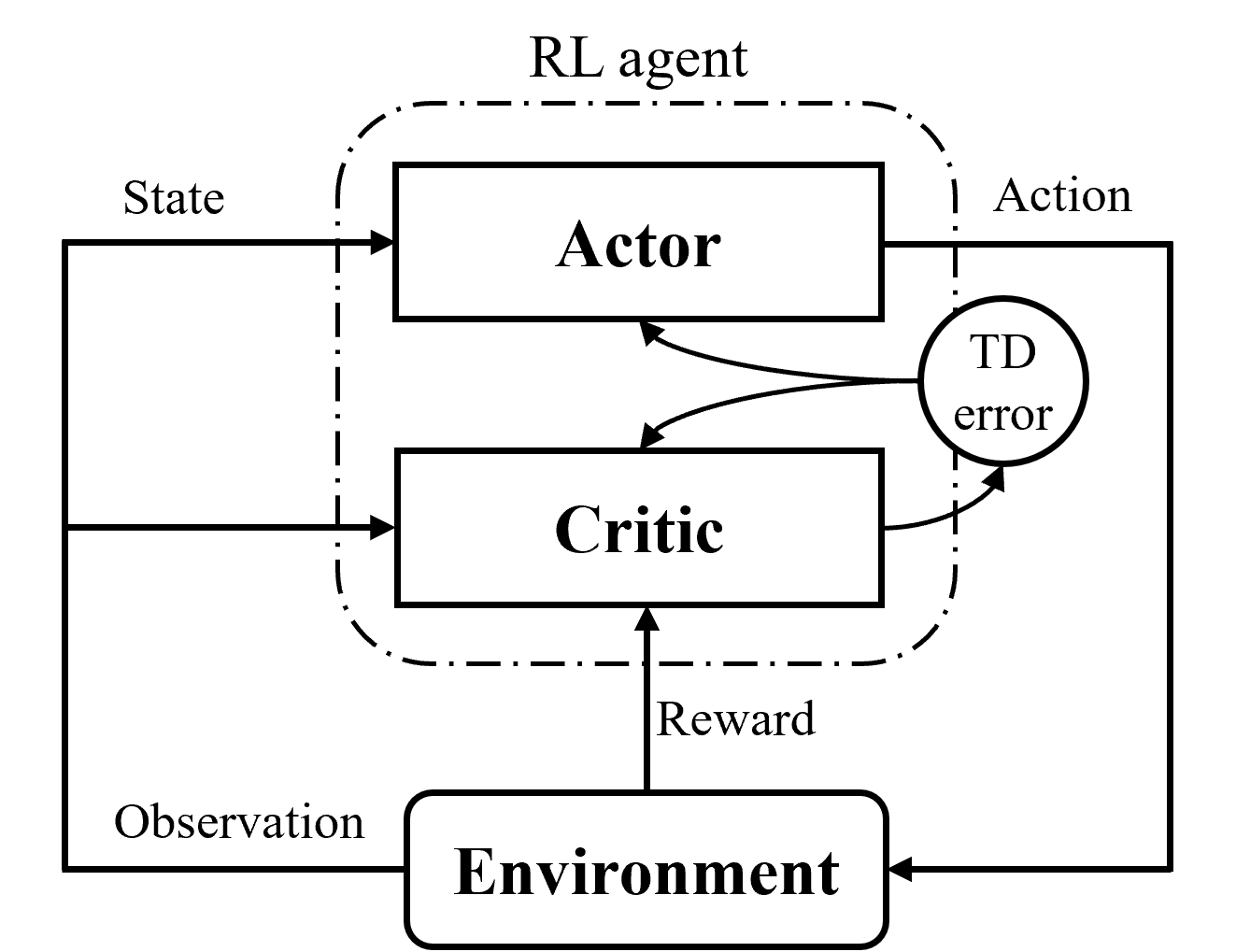}
    \caption{The framework of the AC RL.}
    \label{AC}
\end{figure}

There are various types of RL algorithms, including actor-critic (AC) \cite{konda1999actor}, state-action-reward-state-action (SARSA) \cite{rummery1994line}, and deep RL (DRL) \cite{franccois2018introduction}. AC is an RL technique that integrates an actor, which chooses actions, and a critic, which evaluates the chosen actions, as shown in Fig.~\ref{AC}. The actor updates the policy to maximize rewards, while the critic provides a value estimate to guide the actor's learning process. This synergy allows for efficient policy improvement in decision-making tasks. Similarly, SARSA is a model-free, on-policy RL approach that updates the value function and policy simultaneously. It selects actions based on the current policy, receives rewards, and then updates the policy using the reward and the value of the next state-action pair, ensuring the learning process aligns with the agent's behavior. DRL, on the other hand, integrates NNs with RL to handle high-dimensional data and large state spaces. It enables agents to learn complex behaviors by automatically extracting features from raw inputs, leading to advanced performance in a wide range of tasks from gaming to robotics.

Online learning and RL excel at making online decisions based on received rewards. By appropriately modeling states, actions, and rewards, the wireless transmitter can adaptively fine-tune its transmission parameters. For example, in the context of link adaptation, the states can be represented as channel conditions, actions as the MCS index, and rewards as the data rate. This approach allows us to train the transmitter to optimally adjust the MCS index at each transmission time interval (TTI). Similarly, RL offers a model-free approach to learning optimal power control and user scheduling policies through direct interaction with the environment. This adaptive capability allows RL to dynamically adjust strategies in response to changing channel conditions and user demands, surpassing traditional feedback-based methods. In addition, by utilizing RL, communication and co-designed systems can dynamically make trade-offs between latency and accuracy, efficiently manage limited resources, and adapt to environment changes autonomously. This strategy ensures flexibility and scalability, positioning it as a crucial approach in the evolving landscape of wireless network control systems. 

\subsection{Other Data-driven Methods}
Other data-driven methods~\cite{james2013introduction}, such as hypothesis testing, linear regression, and Bayesian inference, are simpler and more interpretable than advanced techniques like ML, DL, online learning, and RL. They are easier to implement and understand, requiring fewer computations. However, they may lack the flexibility and predictive power of ML and DL, which excel with large, complex datasets and dynamic environments.

The following are some typical algorithms used in traditional data-driven methods. Hypothesis testing evaluates the evidence in a sample to decide if a hypothesis about a population is likely true by comparing a test statistic against a critical value derived from the significance level to determine if the null hypothesis (no effect) can be rejected in favor of the alternative hypothesis (effect exists). Linear regression models the relationship between a dependent variable and one or more independent variables by fitting a best-fit line, estimating coefficients that quantify the impact of independent variables on the dependent variable, and aiming to minimize prediction errors using the least squares method. Bayesian inference updates the probability of a hypothesis based on new evidence by starting with a prior probability, combining this with the likelihood of the observed data to produce a posterior probability, and providing a continuous and dynamic update of beliefs as more data becomes available.

In wireless communication, these traditional data-driven methods play vital roles. Hypothesis testing is used for signal detection and determining the presence of signals amidst noise. Linear regression can model and predict signal strength, interference patterns, and other performance metrics based on environmental factors. Bayesian inference helps in link adaptation by continuously updating the probability of different channel conditions, improving decision-making in varying environments. These methods offer robust, interpretable solutions that complement advanced ML and DL techniques, enhancing the overall reliability and efficiency of wireless communication systems.

%% file: link_adaptation.tex
\section{Link Adaptation}\label{sec3}
Link adaptation is a technique employed in wireless communication systems to enhance transmission performance by adjusting transmission parameters based on channel conditions. This includes dynamically optimizing factors such as MCS index, data rate, and power level to achieve the highest possible reliability and data rate for a given wireless link.
Rate adaptation involves adjusting MCSs to optimize both data rate and reliability, while power adaptation focuses on adjusting transmit power to attain desired signal quality and minimize interference.
This adaptability is crucial in optimizing wireless communication, where channel quality can fluctuate due to variables such as distance, interference, fading, and environmental factors. This section primarily addresses rate adaptation, with further details on power adaptation provided in Section~\ref{sec7}.

\subsection{Traditional Link Adaptation}
\subsubsection{Outer Loop Link Adaptation (OLLA)}
\begin{figure}[t]
    \centering
    \includegraphics[width=1\linewidth]{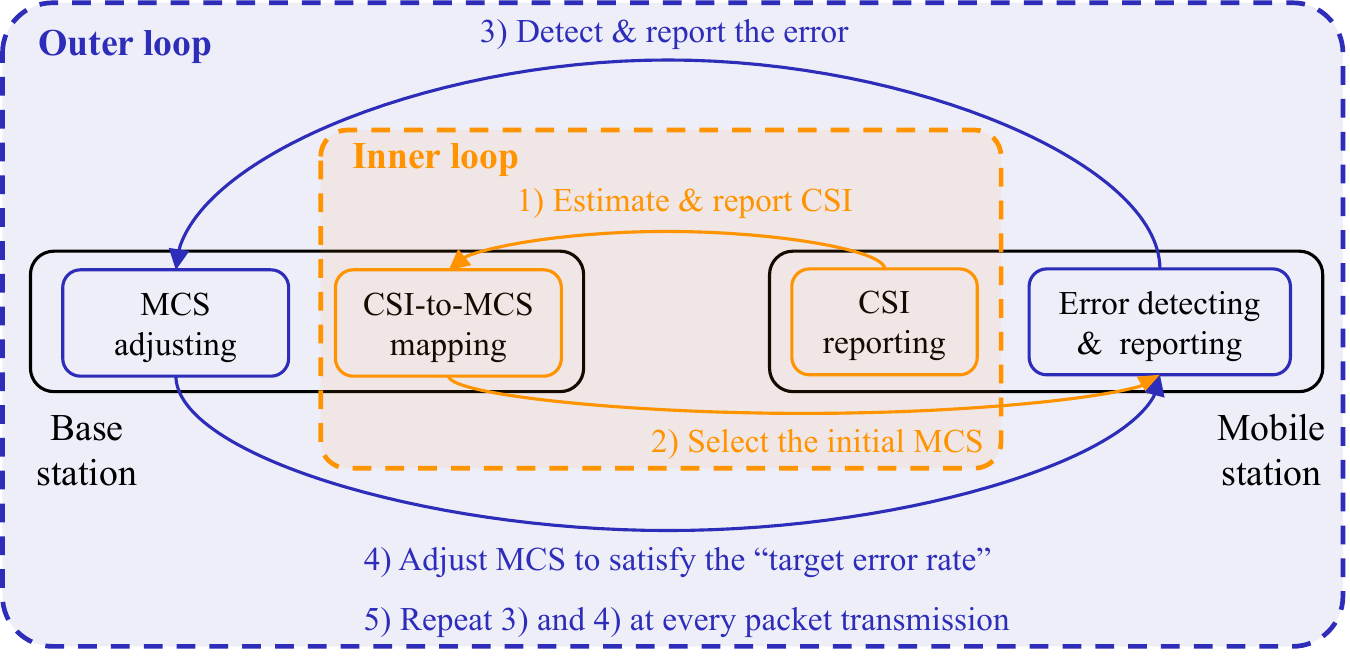}
    \caption{Diagram of ILLA and OLLA}
    \label{fig: ILLA & OLLA}
\end{figure}
In Fig.~\ref{fig: ILLA & OLLA}, the orange region illustrates inner loop link adaptation (ILLA), which uses channel state information (CSI) from user equipment (UE) to estimate the signal-interference-to-noise ratio (SINR). 
ILLA then maps this SINR to an MCS index either to achieve a target average block error rate (BLER) or to maximize throughput via an offline lookup table. 
However, CSI reports are prone to inaccuracies due to reporting delays, detection errors, mapping inaccuracies, and rapid interference, which render ILLA inadequate for highly demanding applications. Consequently, OLLA was developed to address these inaccuracies.
OLLA compensates for CSI reporting errors by dynamically adjusting the SINR threshold based on real-time feedback received from UE, such as acknowledgment (ACK) and non-acknowledgment (NACK) signals. 
This adaptive approach helps OLLA converge the average BLER towards a predefined target, provided the step ratio between upward and downward adjustments is appropriately set.

Nevertheless, in ultra-reliable low latency communications (URLLC) scenarios demanding extremely high reliability, OLLA faces challenges in tuning the step sizes and balancing convergence speed with spectral efficiency loss.
Moreover, traditional OLLA methods may struggle to adapt to the dynamic and rapidly changing channel conditions typical of URLLC environments, potentially leading to suboptimal MCS selections and reduced system performance. Feedback limitations from receivers in URLLC scenarios can exacerbate delays and errors in MCS selection, affecting overall system efficiency.

\subsubsection{Optimization-based Methods}
Mixed integer nonlinear programming (MINLP) is a mathematical optimization framework designed for problems involving both continuous and discrete decision variables. 
It is particularly suited for optimizing MCSs in wireless communication systems, where objectives and constraints often include nonlinear functions.
In MINLP formulations, decision variables can be either continuous (such as power levels) or discrete (like MCS indices and rates). Objectives typically aim to maximize throughput, minimize transmission power, or achieve specific BLER targets. Constraints ensure MCS selections meet criteria such as channel capacity requirements, quality of service (QoS) guarantees (e.g., minimum data rates, maximum bit error rates), and power limits.

However, these problems are computationally intensive due to their non-convex nature, exacerbated by nonlinear functions and integer variables. Consequently, finding optimal solutions often requires heuristic algorithms, and scaling to large scenarios demands substantial computational resources and time, posing challenges for real-time decision-making.
Moreover, MINLP problems are prone to multiple local optima, complicating the search for global solutions that optimize link adaptation performance. The discrete nature of decision variables further adds complexity to the optimization process.

To address the drawbacks of traditional link adaptation methods, numerous studies have explored learning-based approaches that offer greater flexibility and adaptability to changing channel conditions. These methods aim to enhance MCS selection, potentially boosting the efficiency and performance of wireless communication systems. In the subsequent sections, we delve into these innovative approaches.

\subsection{ML for Link Adaptation}
ML offers an alternative to complex theoretical analyses in achieving optimal solutions, such as selecting the optimal transmit mode. This makes classifiers valuable for enhancing the design and optimization of dynamic wireless communication systems.
Yang et al.~\cite{yang2019adaptive} proposed a novel framework integrating supervised ML techniques like KNN and SVM for adaptive spatial modulation-MIMO systems. 
They transformed transmit antenna selection (TAS) and power allocation (PA) problems into data-driven prediction tasks, employing a feature vector generator that considers both modulus and correlation of channel matrix coefficients. 
Their ML-based algorithms are compared with conventional optimization-driven methods such as exhaustive search and max-$d_{\min}$ algorithms in numerical simulations, demonstrating superior BLER performance.
In contrast to Yang et al.~\cite{yang2019adaptive}, which predicted optimal transmission parameters directly, Norolahi et al.~\cite{norolahi2023machine} employed a nonlinear soft-margin SVM to estimate adaptive modulation and coding (AMC) and SINR. These estimates were utilized to adjust modulation type, coding rate, and transmit power across eNode B connections. Simulation results indicated higher success rates, lower power consumption, increased capacity, and simplified implementation compared to conventional approaches.

A detailed comparison of the methods discussed is presented in Table~\ref{tab: ML_LA}.
\begin{table*}[t] 
	\caption{ML approaches for link adaptation.}
 \centering
	\label{tab: ML_LA}
 \scriptsize
	\begin{tabular}[c c c c]{|m{0.1\textwidth}<{\centering}|m{0.2\textwidth}<{\centering}|m{0.2\textwidth}<{\centering} |m{0.3\textwidth}<{\centering}|}
		\hline
		& \multicolumn{2}{m{0.48\textwidth}<{\centering}|}{Yang et al.~\cite{yang2019adaptive}}  &    Norolahi et al.~\cite{norolahi2023machine}  
  \\ \hline
  Link Adaptation Problem & \multicolumn{2}{m{0.48\textwidth}<{\centering}|}{TAS and PA in spatial modulation MIMO systems} & Power control and link adaptation \\ \hline
  ML Method & KNN & $R$-order $2$-class SVM & SVM\\ \hline
Input &  \multicolumn{2}{m{0.48\textwidth}<{\centering}|}{Normalized features regarding modulus and correlation of channel matrix coefficients}  & 
Signals in (orthogonal frequency division multiplexing) OFDM subchannels
\\ \hline
Output & \multicolumn{2}{m{0.48\textwidth}<{\centering}|}{Transmit antenna subsets or power allocation matrix} & Modulation type and SINR
            \\  \hline 
Pros (+) and Cons (-) &  \multicolumn{2}{m{0.48\textwidth}<{\centering}|}{
\begin{itemize} \scriptsize
\item[(+)] Lower complexity compared to conventional methods; improved BLER performance.
\item[(-)] Exponential increase in classes with antenna count; sensitivity to parameter tuning.
\end{itemize}
} & 
\begin{itemize} \scriptsize
\item[(+)] Accurate modulation recognition and SINR estimation; higher AMC success rate.
\item[(-)] Reliance on fixed look-up tables.
\end{itemize}
  \\ 
\hline
\end{tabular}
\end{table*}

\subsection{DL for Link Adaptation}
Yang et al.~\cite{yang2019adaptive} proposed a feed-forward DNN-based multi-label classifier for adaptive spatial modulation MIMO parameter prediction. 
Their neural network includes fully connected layers, ReLU activation, and softmax output, facilitating adaptive operations during training to extract implicit feature information. 
Simulation results demonstrated comparable performance to optimal exhaustive-search-based max-$d_{\min}$ algorithms.
Saxena et al.~\cite{saxena2019contextual} addressed the link adaptation problem as an online stochastic policy optimization using a contextual MAB. 
They employed an NN model to predict transmission success probability for each MCS based on contextual vectors encompassing channel state and other link-side details. Numerical results showed that their approach increases average link throughput by up to $25\%$ compared to OLLA.

In 5G scenarios, enhanced mobile broadband (eMBB) and URLLC cater to vastly different applications. eMBB aims for high throughput for services like train/airplane communications and 4K/8K video streaming. At the same time, URLLC targets ultra-reliability with latency within $1$ ms and $99.999\%$ packet success probability.
Huang et al.~\cite{huang2021deep} proposed DELUXE, which is a DL-based link adaptation method for URLLC multiplexing with eMBB. DELUXE transforms high-dimensional input from eMBB transmissions into a low-dimensional representation to minimize information loss and simplify neural network processing. 
Inputs include eMBB transmission details influencing BLER, such as code block length, URLLC puncturing, and channel conditions. The NN predicts BLER for each MCS, with a calibration mechanism adjusting predictions based on recent decoding results to manage prediction errors. 
Simulation in a link-level 5G NR simulator showed that DELUXE outperforms EESM, a CSI-based link adaptation method, particularly under URLLC puncturing scenarios, ensuring eMBB reliability while maintaining throughput comparable to EESM.
Hussien et al.~\cite{hussien2021towards} introduced a novel approach to link adaptation by developing a multi-label multi-class classification model. 
This model predicts both optimal and suboptimal MCS configurations, aiming to enhance system throughput. 
They devised a unique loss function tailored to mitigate retransmissions stemming from high-rate MCS classification errors, resulting in a noteworthy $50\%$ reduction compared to conventional methods. 
Furthermore, they explored various subdataset selection criteria to evaluate their impact on classification accuracy in link adaptation.

Unlike previous methods that directly predict MCS performance, Mandelli et al.~\cite{mandelli2021troll} proposed training of OLLA (TOLLA), which optimizes OLLA's parameters dynamically for improved performance across diverse scenarios. TOLLA leverages runtime network data, including initial SINR estimates, MCS choice, and transmission outcomes, using binary cross entropy for training. This approach ensures efficient backpropagation with closed-form derivations. Compared to generic ML approaches, TOLLA aligns closely with the structure of the link adaptation problem, offering straightforward implementation and minimal complexity overhead beyond OLLA baselines. They also introduced a CSI correction term to refine incoming SINR estimates and applied gradient detaching to enhance TROLL's initialization. Evaluations against OLLA using 3GPP-calibrated simulations demonstrated that TROLL achieves optimal parameters for URLLC scenarios, yielding superior BLER matching and $10.1\%$ higher average spectral efficiency. This capability obviates the need for exhaustive parameter searches, crucial for applications demanding rapid convergence and stringent reliability targets.

In a pioneering effort on transmit power control, Lee et al.~\cite{Lee} applied CNNs for the first time in wireless communication systems. They proposed a distributed deep power control scheme utilizing only local CSI, aimed at reducing signaling overhead in large-scale user environments. Simulation results indicated that their approach achieves comparable or higher spectral and energy efficiency than conventional weighted minimum mean square error schemes, with significantly reduced computation time.

Table~\ref{tab: DL_LA} presents a detailed comparison of the methods discussed in this review.
\begin{table*}[t] 
	\caption{DL approaches for link adaptation.}
 \centering
	\label{tab: DL_LA}
 \scriptsize
	\begin{tabular}{|m{0.06\textwidth}<{\centering}|m{0.12\textwidth}<{\centering}|m{0.12\textwidth}<{\centering} |m{0.12\textwidth}<{\centering}| m{0.12\textwidth}<{\centering}|m{0.12\textwidth}<{\centering}|m{0.12\textwidth}<{\centering}|}
		\hline
		& Yang et al.~\cite{yang2019adaptive}  &    Huang et al.~\cite{huang2021deep} &  Mandelli et al.~\cite{mandelli2021troll} & Saxena et al.~\cite{saxena2019contextual} & Hussien et al.~\cite{hussien2021towards} &  Lee et al.~\cite{Lee} \\ \hline
  Link Adaptation Problem & TAS and PA in spatial modulation MIMO systems  & MCS selection for eMBB/URLLC multiplexing in 5G NR & MCS selection & MCS selection & MCS selection & Power control in multi-user interference
 \\ \hline
  NN Structure & DNN & DNN & Backpropagation in OLLA & DNN & CNN & CNN \\ \hline
Input & Normalized features regarding modulus and correlation of channel matrix coefficients
 & Low-dimensional representation of eMBB code block size, URLLC puncturing information, and channel condition & Initial SINR estimate, transmission statistics, MCS, received failure indicator & Reported CQI values, long-term average SINR, and receiver speed relative to transmitter & CSI in different selection cases & Approximated channel matrix
    \\  \hline   
    Output & Transmit antenna subsets or power allocation matrix & Predicted $1-\text{BLER}$ for each MCS & Optimized OLLA parameters & Transmission success probability for each MCS & Vector in $\{0, 1\}^{C}$, where $C$ is the number of available transmission modes & Normalized transmit power \\ \hline
    Loss Function & Cross-entropy & Mean square error & Binary cross entropy & Cross-entropy & Cross-entropy with added penalization for false positive predictions & Square of matrix $2$-norm of the difference 
 \\ \hline
    Pros (+) and Cons (-) & \begin{itemize}  \scriptsize
        \item[(+)] Lower complexity than traditional methods; improved BLER performance.
\item[(-)] Class count increases exponentially with antennas.
    \end{itemize}
    & (+) Achieves target eMBB BLER without throughput sacrifice. & 
    \begin{itemize} \scriptsize
        \item[(+)] Simple implementation; minimal complexity atop OLLA; handles BLER constraints efficiently.
    \end{itemize} & 
    (+) Better adaptiveness to varying channels & (+) Predicts all possible labels for successful transmission, reducing retransmissions. & (+) Reduces computation time; captures spatial channel features.
\\ \hline
	\end{tabular}
\end{table*}

\subsection{Online Leaning for Link Adaptation}
To mitigate parameter tuning challenges, various studies have explored online learning algorithms that dynamically adjust policies based on received ACKs to enhance adaptation. Some studies have adapted the exploration process by monitoring changes in ACK statistical characteristics to track variations in the probability of successful transmission. For instance, Combes et al.~\cite{combes2018optimal} focused on optimizing rate adaptation in 802.11 systems by framing it as an online stochastic optimization problem. They introduced the Graphical-Optimal Rate Sampling (G-ORS) algorithm, leveraging graphical unimodality in link adaptation. Graphical unimodality signifies a structural property of throughput functions represented by an undirected graph. Vertices in this graph denote available decisions such as mode and rate pairs in rate adaptation. The adjacency of vertices indicates a relationship in expected throughput between decisions. Crucially, this property facilitates efficient local search algorithms to identify optimal decisions, as paths exist in the graph from any decision to the optimal one with strictly increasing expected throughput. Additionally, they utilized sliding windows and exponential weights to handle non-stationary radio environments. Evaluation of the G-ORS algorithm in an 802.11n test-bed demonstrated superior performance, adaptability to varying channel conditions, and efficiency in maximizing throughput compared to existing schemes like Minstrel HT, Atheros MIMO RA, and MiRA.
In contrast, Lei~\cite{lei2022online} modeled changing channels as a piecewise stationary MAB, assuming observable channel changes occur at a much lower rate than data transmissions. Unlike Combes et al.~\cite{combes2018optimal}, who passively tracked non-stationary channel variations using sliding windows and exponential weights, this work combined TS with a cumulative sum (CUSUM) scheme to actively detect channel changes. Upon detecting a change, learned parameters were reset. Numerical simulations showed that this approach outperformed passive methods like sliding windows in detecting changes. Despite the theoretical superiority of the UCB method over TS, simulations indicated that UCB was prone to early suboptimal decisions, resulting in higher regret compared to TS.
In a recent study, Tong et al.~\cite{tong2023model} proposed a hybrid approach that integrates adaptive discounting with generalized likelihood ratio change detection. This method facilitates simultaneous tracking of slowly varying channels and detection of abrupt changes, specifically tailored for underwater acoustic communications.

The studies by~\cite{combes2018optimal, lei2022online, tong2023model} modeled each transmission parameter as an arm, which can lead to high model complexity. 
Traditional MAB approaches require rewards to be independently and identically distributed, limiting them to static channels. Although they integrate passive or active detection methods, these methods converge slowly in the presence of fast-varying channels. Consequently, some approaches leverage side information about channel states or estimate latent channel states to enhance algorithm adaptability.

Pulliyakode and Kalyani~\cite{pulliyakode2017reinforcement} redefined offset values as bandit arms rather than transmission rates. They introduce the probably approximately correct (PAC) algorithm, surpassing TS and UCB algorithms for OLLA in wireless systems. By exploiting the specific arm structure unique to OLLA, the authors significantly reduce exploration complexity, improving algorithm efficiency and maintaining target BLER levels. Many existing TS and UCB-based approaches focus on maximizing reward without targeting specific BLER levels. In contrast, the PAC algorithm is designed to achieve known target BLERs across the system, crucial for optimizing network capacity and overall performance. TS and UCB algorithms may be less effective for optimizing system-wide metrics compared to individual-focused metrics.

To address OLLA limitations, Saxena and Jald{\'e}n~\cite{saxena2020bayesian} proposed an updating strategy with the Bayesian link adaptation scheme, called BayesLA. 
They treat each combination of estimated CSI and MCS candidates as an arm, transforming the non-stationary MAB problem into a stationary one. Simulations using a fourth-generation (4G) wireless link model over a Rayleigh fading channel demonstrate that both BayesLA and OLLA maintain BLER close to a target of 0.1. However, BayesLA exhibits superior BLER performance over time compared to OLLA, achieving higher throughput under a BLER target of 0.3. This optimized performance is ideal for simulated wireless scenarios.

Recently, reinforcement learning link adaptation (RLLA) has been proposed to automate MCS behavior learning and directly optimize link performance objectives. However, RLLA may require higher resource allocation and exhibit slower adaptation to channel variations. To overcome these limitations, Saxena et al.~\cite{saxena2021reinforcement} extended the TS approach with latent TS (LTS). LTS encodes the environment using a low-dimensional latent state model to probabilistically model channel SINR and refines this model based on feedback received after each transmission. They employed an offline link model mapping channel estimates to MCS performance levels.
LTS extends its applicability to fading wireless channels via a tuning-free mechanism that automatically tracks channel dynamics. By smoothing the SINR probability distribution with a suitable function, LTS adapts to fading channels without manual parameter adjustment, thereby enhancing robustness and adaptability. Numerical evaluations demonstrate LTS's faster convergence, higher throughput performance, adaptability to channel fading, lower complexity, and stable steady-state performance compared to OLLA and existing RLLA methods~\cite{gupta2018low, combes2018optimal}. These findings underscore LTS's effectiveness in enhancing link performance and robustness in wireless networks, positioning it as a promising algorithm for efficient and tuning-free link adaptation.

Praveen et al.~\cite{praveen2021reinforcement} proposed a contextual MAB approach that speeds up link adaptation without waiting for ACK/NACK feedback from the UE. Their context vector incorporates UE velocity, carrier frequency, and long-term average SINR of the link. Simulation results highlight the effectiveness of this approach in addressing challenges posed by outdated CQI feedback and fast-fading channel conditions in URLLC scenarios. The method demonstrates superior performance in reliability, spectral efficiency, and adaptability compared to traditional OLLA techniques, presenting it as a promising solution for optimizing link adaptation in 5G networks.
Finally, UCB and TS are well-known methods for unconstrained MAB problems. However, their optimality may diminish under structural constraints on arms. Dey et al.~\cite{dey2023linearly} modeled closed-loop link adaptation as a linearly constrained stochastic MAB problem and propose an algorithm that leverages model structure to minimize the search space for optimal actions during packet transmission windows. 

A detailed comparison of these methods is provided in Table~\ref{tab: OL_LA}.

\begin{table*}[t] 
	\caption{Online learning approaches for link adaptation.}
 \centering
	\label{tab: OL_LA}
 \scriptsize
	\begin{tabular}[c c c c c ]{|m{0.1\textwidth}<{\centering}|m{0.15\textwidth}<{\centering}|m{0.2\textwidth}<{\centering} |m{0.2\textwidth}<{\centering}| m{0.1\textwidth}<{\centering}|}
		\hline
		 Works  & Link Adaptation Problem & Online Learning Methods & Adaptive Methods for Varying Channels & Target BLER consideration  \\ \hline
   Combes et al.~\cite{combes2018optimal} & Rate adaption in 802.11 systems & Unimodal stochastic MAB (Arm: mode and rate pairs) & Sliding windows and exponential weights & $\times$  \\ \hline
   Lei~\cite{lei2022online} &  Rate adaption & Piecewise stationary MAB (Arm: rate) & Change detection with CUSUM & $\times$ \\ \hline
   Tong et al.~\cite{tong2023model} & Frequency and rate adaptation in underwater acoustic communications & Stochastic MAB (Arm: rate and frequency pairs) & Adaptive discounting with generalized likelihood ratio change detection & $\times$ \\ \hline
   Pulliyakode and Kalyani~\cite{pulliyakode2017reinforcement} & MCS selection & Stochastic MAB (Arm: offset values in OLLA) & Adjustment of offsets based on ACK/NACK feedback  & 0.1 \\ \hline
   Saxena and Jald{\'e}n~\cite{saxena2020bayesian} & MCS selection & Contextual MAB (Arm: CQI and MCS pairs) & Arm updates based on estimated CQI & $0.1$ \\ \hline
  Saxena et al.~\cite{saxena2021reinforcement} & MCS selection in fading channels & Latent bandit (latent state: SINR) & SINR PDF relaxation through convolution with Normal distribution  & $\times$ \\ \hline 
  Dey et al.~\cite{dey2023linearly}& MCS selection & Linearly constrained error MAB (Arm: encoder) & Adjustment of encoder selections based on error-free transmissions & $\checkmark$ \\ \hline
	\end{tabular}
\end{table*}

\subsection{RL for Link Adaptation} 
DRL offers a notable advantage in adapting effectively to environmental variations without prior knowledge.
Kela et al.~\cite{kela2022reinforcement} proposed an RL-based OLLA algorithm for uplink scenarios aimed at optimizing spectral efficiency for high-reliability applications while respecting the packet delay budget (PDB). This novel approach provides a perspective for delay-sensitive traffic to leverage ML models capable of efficiently handling strict PDB and reliability constraints, thus preventing system overload due to excessive robustness.
Wu \cite{wu2020q} introduced a Q-learning framework that integrates the strengths of ordinary link adaptation (OLA) and reinforcement-based methods, incorporating historical hybrid automatic repeat request (HARQ) data and CSI feedback within its state space. This framework ensures seamless transitions between MCSs and transmission ranks. Moreover, leveraging a neural network for approximating the state-action function enhances the efficiency and efficacy of the link adaptation process.
Makridis \cite{makridis2020reinforcement} proposed a novel framework utilizing RL algorithms such as DQN and Proximal Policy Optimization (PPO) to optimize downlink link adaptation in 5G networks. This approach enables agents to dynamically adjust AMC parameters based on real-time network conditions, eliminating the reliance on predefined BLER targets and facilitating more flexible and adaptive link adaptation schemes. This adaptability allows autonomous adjustment to varying channel conditions without manual intervention, thereby enhancing overall system performance.

In addition to MCS selection, Xu et al.~\cite{xu2020augmenting} presented an RL-based framework for efficient rate adaptation in drive-thru Internet scenarios. This framework utilizes historical channel measurements and transmission feedback to train a NN for optimal rate selection in vehicle uplink data transmissions. It addresses transmission failures caused by channel fading and medium access collisions, demonstrating superior performance over traditional rate adaptation schemes across various levels of environmental challenges.
Ye et al.~\cite{ye2023deep} addressed the challenge of outdated CSI feedback by proposing a DRL-based link adaptation technique. Their approach incorporates various sources of information into the state representation, including the most recent CSI from the UE, the CSI difference over consecutive TTIs, and historical data. This comprehensive state enables the joint decision-making process for MCS selection.
The agent receives immediate rewards based on transmission outcomes: it receives the effective data rate for successful transmissions and -1 for failures. To balance throughput and BLER, they introduced a classified experience replay (CER) mechanism. CER partitions experience into successful and failed categories, from which a fixed proportion is sampled for replay. Adjusting these proportions allows for diverse training experiences.
Moreover, to mitigate the impact of frequent MCS switching on signaling and system overheads, they proposed a switching-controlled $\epsilon$-greedy strategy. This strategy optimizes the trade-off between transmission quality and switching costs.

The detailed comparison among the mentioned methods is shown in Table~\ref{tab: RL_LA_new}.
\begin{table*}[t] 
	\caption{RL approaches for link adaptation.}
 \centering
	\label{tab: RL_LA_new}
 \scriptsize
	\begin{tabular}[c c c c]{|m{0.08\textwidth}<{\centering}|m{0.15\textwidth}<{\centering}|m{0.15\textwidth}<{\centering} |m{0.15\textwidth}<{\centering}| m{0.15\textwidth}<{\centering}|m{0.15\textwidth}<{\centering}|}
		\hline
		& Ye et al.~\cite{ye2023deep}  & Kela et al. \cite{kela2022reinforcement} & Wu \cite{wu2020q} & Makridis \cite{makridis2020reinforcement} & Xu et al.~\cite{xu2020augmenting} \\ \hline
 State & Historical CSIs, CSI changes between consecutive TTIs, action decisions & Offset values for OLLA status & Previous and current CQI, rank indicator, MCS index, and ACK/NACK & CQI index variations, HARQ-ACK averages, backoff values & Recent SNR records \\ \hline
Action & MCS index selection & Increment or decrement of OLLA offset & Layer and MCS selection & Adjustment of offset value steps & Rate adaptation \\ \hline
Rewards & Effective data rate for ACK, $-1$ for NACK &  Ratio of received data bytes to total resource block usage + delay penalty for failures & TBS of previous transmission for ACK & Average downlink throughput  & Amount of data received after transmission attempt \\ \hline
RL Methods & Deep Q learning with classified experience reply
 & Q-learning & Deep Q-learning & Deep Q-learning
 & Deep Q-learning
\\ \hline
    Pros (+) and Cons (-) & (+) Balances throughput and BLER trade-offs
    & (+) Minimize radio resource use and avoids delay violations.
 &  \begin{itemize} \scriptsize
        \item[(+)] Outperforms OLA across metrics
        \item[(-)] Requires more iterations to converge
    \end{itemize} & (+) Higher throughput; no dependency on predefined fixed parameters;  maintains the minimal radio network intervention & (+) Outperforms traditional schemes, robust to different channels and contention levels \\ \hline
	\end{tabular}
\end{table*}

\subsection{Summary and Comparisons}
In conclusion, learning-based link adaptation methods can be categorized into two distinct approaches, each offering unique advantages:
\begin{itemize}
    \item \textbf{Direct Prediction Models:}
    These methods train models to directly predict transmission performance based on channel states, leveraging DNNs. By bypassing lookup tables, they exhibit faster adaptation to varying channel conditions compared to OLLA.
    \item \textbf{Offset training within OLLA:}
    This approach integrates learning within the OLLA framework, specifically focusing on training the offset parameter. Determining the optimal offset is generally challenging, but this method addresses it through a learning approach. Unlike the black-box nature of direct prediction models, this method provides more interpretability since it operates within the OLLA framework, ensuring adherence to BLER constraints.
\end{itemize}

ML and DL excel in capturing intricate nonlinear patterns within data, enabling automated feature extraction without explicit modeling or engineering. This adaptability is particularly advantageous for optimizing MCS selection based on complex channel conditions. However, effective utilization hinges on abundant labeled training data, which can be arduous to collect and annotate, especially for niche or rare channel scenarios. The quality and availability of such datasets significantly influence the performance and generalizability of ML and DL models. Moreover, the computational demands of training and deploying these models can be prohibitive, especially in resource-constrained wireless environments, limiting their real-time applicability.

Online learning intelligently balances between exploration (testing different MCS options) and exploitation (leveraging the best-known option). This approach gathers performance data across MCS choices while maximizing the use of the most effective scheme. However, the inherent trade-off between exploration and exploitation can occasionally lead to sub-optimal decisions, as the algorithm explores less-performing MCS options, potentially reducing overall system performance.

RL offers adaptive decision-making in dynamic wireless environments, adjusting to varying channel conditions, interference levels, and user demands. RL optimizes long-term objectives such as throughput or spectral efficiency through iterative learning from interactions with the environment. While versatile, RL training processes are computationally intensive and time-consuming, requiring multiple iterations to refine policies. Efficient training strategies and hardware optimizations are essential for scaling RL-based adaptations, especially when managing extensive state and action spaces in wireless networks.

\subsection{Challenges and Open Issues}
Despite the progress of data-driven link adaptation, several challenges persist in academic and industry settings for next-generation wireless communication networks:
\begin{itemize}
    \item \textbf{Adaptiveness to heterogeneous networks:}
    Mobile devices frequently switch between Wi-Fi, cellular, and Bluetooth networks, requiring seamless link adaptation. However, data-driven methods may need time to adjust when transitioning between different network types, potentially affecting performance. Improving the generalization of data-driven approaches, reducing complexity, and ensuring rapid convergence in real-world applications are critical.
    
    \item \textbf{Extremely high reliability requirements:} 
    Existing data-driven link adaptation algorithms often focus on maximizing average throughput without considering stringent reliability demands, such as $\text{BLER} = 10^{-5}$. Many algorithms evaluate performance at higher BLER targets, like $0.1$, highlighting a need for research to achieve both high throughput and extreme reliability in data-driven link adaptation.

    
    \item \textbf{Hardware and computational constraints:}
    Deploying data-driven link adaptation algorithms, including DL models, on devices with limited computational resources is challenging. Lightweight approaches with low complexity and energy consumption are essential, particularly for wearable devices reliant on efficient transmission parameter adaptation.

    \item \textbf{Interference in dense networks:} 
    Current data-driven link adaptation methods typically address single-link scenarios, neglecting interference from neighboring users in dense networks. Adopting multi-agent learning techniques can enhance link adaptation strategies to handle interdependencies effectively in dense network environments.
\end{itemize}

%% file: user_schedule.tex
\section{User Scheduling}\label{sec4}

User scheduling refers to the process of efficiently selecting a subset of active users from the available user pool and allocating signal channels appropriately based on the specific requirements and priority of each user. In today's world, more and more electronic devices are connecting to networks, and the demands for high-speed data transmission in various network applications are also increasing. Consequently, these pose significant challenges to the efficiency and effectiveness of resource allocation. 

\subsection{Traditional User Scheduling}

\subsubsection{Background}

In a large and complex network, it is challenging to find the optimal scheduling strategy accurately under various constraints and a massive volume of data. In addition, the precise CSI is usually hard to obtain due to complex network structures and stochastic properties of wireless channel conditions. To address these issues, apart from the traditional algorithms that approximate system responses and CSI, model-free methods become powerful tools for accuracy improvement. Another challenge to be concerned about is the rapidly changing system states such as the number of users accessing the network, their demands, and the performance of accessing devices in public networks. To solve this issue, practical user scheduling algorithms are required to have high responsiveness to adapt to dynamic conditions. Additionally, it’s important to ensure fairness of policies while efficiently allocating resources. In general, fairness and efficiency can be trade-offs in communication services considering the inherent allocation constraints and limited resources within a system. 

\subsubsection{Optimization-based Methods}

In order to enhance the performance of communication systems, SINR is a frequently used metric for evaluation, which represents the ratio between the signal power and the interference with noise.  By utilizing SINR or the total information rate as the objective function for optimization and maximizing it, the optimal scheduling results can be obtained based on the current CSI. In addition, other limitations or optimization targets that arise from system design can be similarly included in the objective functions or treated as constraints in the optimization problem. When SINR is increasing, the impact of co-channel interference (CCI) and system noise on communication will be smaller, leading to better system performance. Moreover, the introduction of constraints can pose difficulties in solving this optimization problem, which leads to the fact that the traditional user scheduling algorithms essentially address the challenges of solving optimization problems while maintaining computational speed and efficiency in real-time. 

\subsubsection{Motivation}

In communication systems with different settings, environments, and capacities, the prior objectives attached to these aspects may vary accordingly. Therefore, it is essential for the user scheduling algorithms to exhibit flexibility and scalability to adapt to the diverse objectives associated with different settings. Hence, data-driven approaches hold immense potential in addressing user scheduling problems. By leveraging an abundance of data, ML-based classification methods and DNNs can serve as model-free tools to estimate CSI more accurately and offer valuable insights for subsequent resource scheduling. RL methods can enhance the performance of scheduling policies through iterative processes, continuously optimizing policies based on the evolving system states. These approaches effectively overcome the limitations associated with the absence of system models and provide unified algorithms capable of addressing diverse optimization objectives.

\subsection{ML for User Scheduling}

\begin{figure}
    \centering
    \captionsetup{justification=centering}
    \includegraphics[scale=0.35]{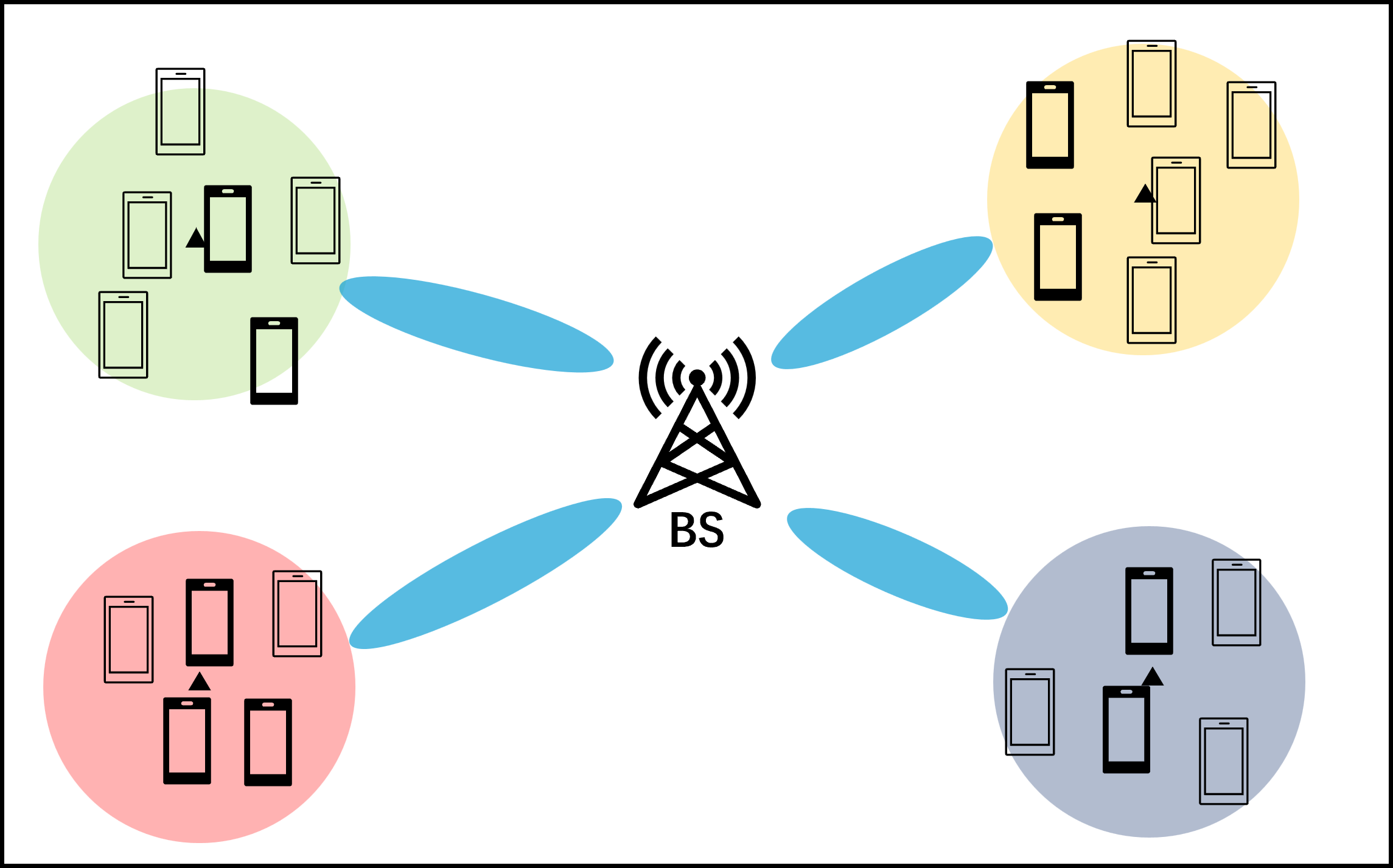}
    \caption{The KM diagram of user clustering.}
    \label{UC}
\end{figure}

A dual-step user scheduling approach for large-scale multi-antenna communication systems was proposed by Feres et al. \cite{feres2022unsupervised}. This approach consists of two steps, where the first step is to classify the CSI of users and cluster them based on the CSI similarity. A transform method called the Grassmannian manifold was used as the metric evaluating similarity, with which the authors implemented KM as the unsupervised clustering approach. In the second step, the users are reallocated into resource-sharing groups to reduce CCI, aiming to increase the SINR of the overall system. Similarly, KM was used by Cui et al. \cite{cui2018unsupervised} to do user clustering based on Euclidean distance for reducing the computational complexity of user scheduling, which is shown in Fig. \ref{UC}. 

SVM was used for distinguishing extreme cases of user scheduling by Cao et al. \cite{cao2018machine}. Their user scheduling method assigns links with maximum or zero energy based on separation results of SVM at the first step, thereby significantly reducing the computation time of the entire user scheduling algorithm.
\begin{center}
\begin{table}
\caption{ML approaches for user scheduling.}
\scriptsize
\begin{tabularx}{0.496\textwidth} { 
  | >{\centering\arraybackslash}m{0.07\textwidth} 
  | >{\centering\arraybackslash}m{0.11\textwidth}
  | >{\centering\arraybackslash}m{0.11\textwidth}
  | >{\centering\arraybackslash}m{0.11\textwidth}| }
 \hline
 & Feres et al. \cite{feres2022unsupervised} & Cui et al. \cite{cui2018unsupervised} & Cao et al. \cite{cao2018machine} \\
\hline
ML Method & K-means & K-means & SVM \\
\hline
Purpose & User clustering & User clustering & Solving extreme cases \\ 
\hline
Training Dataset & \(--\) & \(--\) & Offline optimization results \\ 
\hline
Metrics & CSI similarity & User position & \(--\) \\
\hline
Transform / Kernel & Grassmannian manifold & Euclidean distance & Linear kernel \\
\hline
Complexity & \(\mathcal{O}(tKMN)\)  & \(\mathcal{O}(tKMN)\) &\(\mathcal{O}(NL)\) \\
\hline
Pros(+) or Cons(-) & (+) Using the Grassmannian manifold can represent the similarity of complex vectors. (-) Performance highly relies on initial means and parameter \(K\). & (+) Computation cost is low. (-) Only position similarity is considered. & (+) SVM significantly reduces computational complexity. (-) Scalability is limited and the performance of the system is sacrificed.\\
\hline
\end{tabularx}
\end{table}
\end{center}

\subsection{DL for User Scheduling}

As a representative supervised learning method, DL requires data generated from high-quality policy or optimization solutions to do training. This poses challenges on collecting high-quality data, tuning hyper-parameters and guaranteeing robustness of performance. Therefore, on one hand, DNN can be leveraged to learn from well-designed optimization solutions, as the online computational efficiency of DNN surpasses that of solving complex optimization problems. On the other hand, another idea involves integrating DL with optimization algorithms to speed up the online computation of optimization problems, while preserving the interpretability of the designed user scheduling algorithms.

As wireless networks can be naturally represented as graphs, where nodes represent users or base stations (BSs), while edges represent the communication links within the network. GNNs are well-suited to handle user scheduling problems in complex communication systems. In addition, GNNs are adaptive to dynamic changes in the network topology, which shows great potential in solving user scheduling issues due to user mobility and varying channel conditions.

He et al. \cite{he2022joint} proposed an optimization-based joint user scheduling algorithm using GNNs. The authors first designed a fully optimization-based user scheduling method to address the scheduling problem within a single BS, which utilized successive convex approximation (SCA) to transform non-convexity into a tractable problem. Based on this algorithm, the problem was further formulated into a joint user scheduling problem with the objective of maximizing the signal rate while minimizing the energy required by the BSs. Considering it’s difficult to obtain the prior topological information of the complex wireless communication networks, GNNs are integrated for learning the existing user scheduling strategy in the form of graph representation. From the simulations, the proposed approach achieves close performance to that of pure optimization algorithms, while reducing computational consumption by \(90\%\). Another user scheduling method using GNNs was designed by Zhang et al. \cite{zhang2022learning}, which focuses on the user scheduling of a reconfigurable intelligent surface-assisted multiuser downlink network.

\begin{figure}
    \centering
    \captionsetup{justification=centering}
    \includegraphics[scale=0.47]{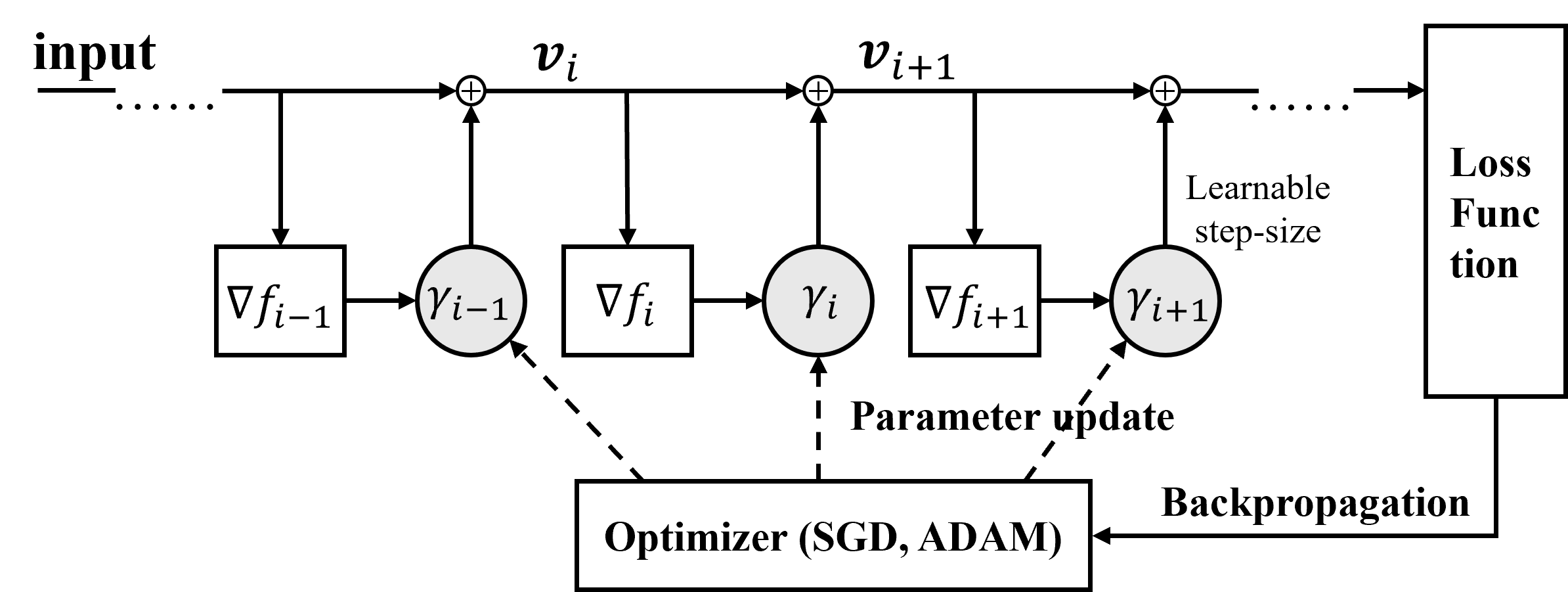}
    \caption{The framework of the deep unfolding technique.}
    \label{DU}
\end{figure}

Another user scheduling approach using the deep unfolding technique was proposed by Xu et al. \cite{xu2023joint}, aiming at solving user scheduling problems in cloud radio access networks (C-RANs) with multiple BSs equipped with a fixed number of antennas. In a dense C-RAN, as BSs can jointly transmit data to scheduled users, the coordination becomes much more complex. The optimization problem was formulated as a weighted minimum mean square error (WMMSE) problem in this paper to maximize the total information rate while minimizing the allocated BS number of each scheduled user. Due to the existence of a non-convex function, a coordinate descent method was used for decomposing the optimization problem into multiple iterative computations, aiming to progressively approach the optimal solutions. As shown in Fig.~\ref{DU}, the step-size \(\gamma_i\) of each iteration step (\(\boldsymbol{v}_{i+1}=\boldsymbol{v}_i-\gamma_i \nabla f_i\)) is learnable by considering the whole optimization process as a DNN, where \(\boldsymbol{v}_{i}\) is the scheduling vector and \(\nabla f_i\) is the derivative of objective function at \(i\) step. From simulation results, the proposed algorithm converges much faster compared to original optimization algorithms with fixed step size. Therefore, the proposed method can provide better scheduling performance and save more computation resources within limited computational time in comparison with other WMMSE-based user scheduling algorithms.

\begin{center}
\begin{table}
\caption{DL approaches for user scheduling.}
\scriptsize
\begin{tabularx}{0.496\textwidth} { 
  | >{\centering\arraybackslash}m{0.07\textwidth} 
  | >{\centering\arraybackslash}m{0.11\textwidth}
  | >{\centering\arraybackslash}m{0.11\textwidth}
  | >{\centering\arraybackslash}m{0.11\textwidth}| }
 \hline
 & He et al. \cite{he2022joint} & Zhang et al. \cite{zhang2022learning} & Xu et al. \cite{xu2023joint} \\
\hline
DL Method & GNN & GNN & Deep unfolding \\
\hline
Purpose & Obtain topology information & Estimate CSI  and solve optimization problem & Accelerate optimization \\ 
\hline
Training Dataset & Sampling and training backward & Sampling and training backward & Pre-run results \\ 
\hline
Input & Scheduled user state, transmit power, graph & Received pilots, user weights & \(--\) \\
\hline
Output & Scheduled user state*, transmit power* & Estimate CSI, BS beamforming matrix & \(--\) \\
\hline
Pros(+) or Cons(-) & (+) DL significantly increases computational efficiency. & (+) This method separates the network into three stages to decouple the complexity. (-) There may exist ineffective solutions.& (+) DL is only used for accelerating computations while preserving the interpretability of the original problem.\\
\hline
\end{tabularx}
\end{table}
\end{center}

\subsection{RL for User Scheduling}

Wei et al. \cite{wei2017user} proposed a policy-gradient-based AC algorithm, which is a model-free DRL approach, to maximize energy efficiency in heterogeneous networks (HetNets). A HetNet consists of a traditional central macro base station (MBS) and several micro-cell or small-cell base stations (SBSs) deployed surrounding the MBS. In comparison to existing literature, this paper presented a novel contribution by introducing the concept of renewable energy. Notably, the paper focused on the hybrid utilization of renewable and traditional energy sources in SBSs, which leads to additional complexities in the optimization of energy efficiency. This scheduling problem with continuous state and action space can be formulated as a discrete-time Markov decision process (MDP) by defining state transition probability as the integration of the state transition probability density function. The action of the policy controls the allocated number of users on each BS, the subchannel index, and the transmission power toward each user. In addition, the reward function in this problem is defined as the ratio of the total information rate over the total conventional power consumption. However, with this definition of rewards, fairness among users is neglected as only the optimization of efficiency is considered when increasing the overall rewards of each trajectory. 

Both the value function and the policy are approximated by neural networks due to the infinite states and action space of the communication system. The AC training process simultaneously updates the value function and optimizes the policy. Moreover, to enhance convergence and reduce variance during the training process, some methods such as the advantage function, and eligibility trace were implemented in this paper.

Another user scheduling approach was proposed by Sharifi et al. \cite{sharifi2023deep}, using deep SARSA reinforcement learning (DSRL) to coordinate service between high-altitude platform stations (HAPSs) and terrestrial base stations (TBSs). The challenge addressed in this paper was that only outdated CSI of the TBSs could be obtained for implementing user scheduling. Specifically, the communication system comprises a HAPS and a TBS, where the HAPS, equipped with N antennas, serves as a backup for the TBS in downlink data transmission. The DSRL algorithm utilizes a reward function defined as the total sum rate of the communication system, while the action function determines the user set assigned to the HAPS. Digging more on the DSRL, on one hand, the potential optimal policy is obtained through the \( \epsilon\)-greedy algorithm. On the other hand, to better acquire system information, this method approximates the model of path loss, considering geometric attenuation, Doppler frequency, and shadow fading. By comparing with three user scheduling methods: a heuristic optimization algorithm, a deep Q-learning (DQL) method, and a random selection method, the proposed DSRL outperforms the other methods when outdated CSI is given, and the degree of advantage improves with the increase of channel estimation error.

Different from centralized learning-based methods, federated learning (FL), as a distributed ML paradigm, can reduce data transmission with the increase of BSs in communication systems, and better protect user privacy and information security. Wu et al. \cite{wu2023adaptive} combined the FL networks with the AC-based DRL to achieve an adaptive user scheduling approach. The intelligent agent trained through this RL method not only aims at maximizing the total sum rate but also seeks to minimize communication loss and time in one global learning round. This algorithm also guarantees the FL convergence, which can be considered a trade-off issue. The simulation results indicate substantial reductions in communication costs without sacrificing the training efficiency of FL compared to the baseline schemes.
\begin{center}
\begin{table}
\caption{RL approaches for user scheduling.}
\scriptsize
\begin{tabularx}{0.496\textwidth} { 
  | >{\centering\arraybackslash}m{0.07\textwidth} 
  | >{\centering\arraybackslash}m{0.11\textwidth}
  | >{\centering\arraybackslash}m{0.11\textwidth}
  | >{\centering\arraybackslash}m{0.11\textwidth}| }
 \hline
 & Wei et al. \cite{wei2017user} & Sharifi et al. \cite{sharifi2023deep} & Wu et al. \cite{wu2023adaptive} \\
\hline
RL Method & AC RL & Deep SARSA & AC RL \\
\hline
Network System & HetNets & HAPSs & FL networks \\
\hline
DNN Approximation & Actor and critic functions & Q-function & Actor and critic functions\\
\hline
State & SINR of each user and battery energy level of each BS & CSI and the last action & CSI, model accuracy and iteration index \\ 
\hline
Action & Scheduled user, subchannel allocation and power allocation & Scheduled user of each station & Activated user ratio, scheduled user and resource block allocation \\
\hline
Reward & SINR and power efficiency & SINR of the whole system & SINR + power consumption \\
\hline
Pros(+) or Cons(-) & (+) Some tricks are used to help accelerate the converge. \newline (-) Trade-off between quality and energy efficiency can not guarantee all users' experience.
& (+) The method works under outdated CSI, showing better robustness in practice. & (+) The method maximizes the reward while ensuring the FL convergence.\\
\hline
\end{tabularx}
\end{table}
\end{center}
\subsection{Challenges and Open Issues}
Although data-driven methods for user scheduling have great potential and performance compared to traditional methods, there are still issues that need to be addressed.
\begin{itemize}
    \item \textbf{Trade-off between overall performance and fairness:} In the above papers, the majority of optimization objectives are to maximize the overall SINR in the system while minimizing energy consumption. However, such objectives fail to address the issue of scheduling fairness, meaning that there might be cases where the performance of service for a few users is sacrificed to maximize the overall system performance. Therefore, finding a balance between performance and fairness in scheduling is still an ongoing direction for improvement in user scheduling algorithms.

    \item \textbf{Variation of user quantity:} While designing scheduling methods, the use of ML, DNN, or function approximation in RL is aimed at addressing scheduling challenges with a large number of users and complex scenarios. However, these methods may lead to computational resource waste when dealing with a small number of users. Therefore, it's important to put improving adaptability to different user quantities into consideration in future scheduling algorithm designs.
\end{itemize}

%% file: spectrum_allocation.tex
\section{Spectrum Allocation}\label{sec5}

In wireless networks, resource allocation is a critical challenge. Such networks often utilize techniques like OFDMA to divide the spectrum and assign portions to different users. The goal is to enable intelligent resource management -- distributing frequency, power, and other factors to minimize overall power consumption while meeting individual user data rate requirements or maximizing total throughput.

A key issue in these systems is spectrum inefficiency. Despite limited utilization, there remains substantial unmet demand. To address this, dynamic spectrum access networks have been developed, leveraging cognitive radio networks (CRNs). These allow secondary users (SUs) to access the spectrum allocated to higher-priority primary users (PUs), as long as PU transmissions are not disrupted. The challenge is allocating spectrum to SUs in a way that maximizes efficiency, avoids interfering with PUs, and satisfies the QoS needs of both user types.

Tackling this resource allocation problem has two main complexities. First, resources can be assigned across multiple orthogonal dimensions - frequency, time, space, coding, and antenna directionality - allowing dynamic, adaptive utilization. Second, there are often conflicting objectives to optimize, such as power, spectrum usage, throughput, and interference management. Some may be hard constraints rather than direct optimization targets. Factors like PU interference thresholds, channel conditions, and user locations must be considered in the optimization process.

The major methods for resource allocation in wireless networks include RL and genetic algorithms, along with other optimization techniques such as game theory, mathematical programming, and heuristic algorithms, which play a crucial role in addressing the resource allocation challenges in wireless networks.
A proper spectrum allocation can lead to higher network capacity in communication networks. However, in the context of CRNs, where spectrum resources are limited, the efficient utilization of the available spectrum becomes crucial. In such networks, it is important to strike a balance between maximizing network capacity and achieving spectral efficiency, considering the conflicting objectives involved.

\subsection{Online Optimization for Spectrum Allocation}
Spectrum allocation in CRNs has been extensively studied using various optimization techniques. Game theory remains a popular method, as demonstrated by \cite{niyato2007game}, who proposed a game-theoretic approach to dynamic spectrum sharing, enabling users to adjust their strategies based on others' actions to reach an equilibrium, thereby reducing interference and improving spectrum utilization. Linear programming (LP) has also been applied by \cite{yousefvand2012interference}, focusing on minimizing interference while meeting the QoS requirements of secondary users through a model that efficiently solves the problem using linear constraints and objectives. To capture the complexity of real-world resource allocation more accurately, \cite{yousefvand2015maximizing} introduced a mixed-integer linear programming (MILP) model that considers both binary decisions and continuous variables.

Convex optimization techniques were explored by \cite{li2015power} for power control, formulating the problem to minimize total transmitted power while meeting the SINR requirements of SUs, thus enhancing energy efficiency. Similarly, the authors in \cite{kai2019joint} developed an integer programming model for joint channel assignment and power allocation to maximize the network's sum rate. In \cite{lin2010optimal}, the authors utilized nonlinear programming to address the dynamic spectrum access problem, focusing on maximizing network utility by accounting for the nonlinear relationship between user utility and spectrum allocation.

Heuristic algorithms, including genetic algorithms and simulated annealing, were investigated by \cite{girgis2014solving} to provide near-optimal solutions with lower computational complexity, suitable for large-scale CRNs. In \cite{mawatwal2021state}, the authors explored distributed optimization techniques, allowing users to optimize their resource usage based on local information, which enhances scalability and robustness in decentralized networks. In \cite{sun2021spectrum}, the authors proposed cooperative optimization strategies, where users share information and coordinate actions to improve overall network performance, leading to better spectrum utilization and reduced interference.

In \cite{zou2008qos}, the authors emphasized the importance of QoS-aware spectrum allocation, ensuring that the allocated spectrum meets the QoS requirements of various applications, and balancing efficiency with service quality. Lastly, \cite{alonso2021multi} introduced multi-objective optimization techniques that consider conflicting objectives such as maximizing throughput while minimizing interference, providing a set of Pareto optimal solutions for trade-offs between different performance metrics.

\subsection{RL for Spectrum Allocation }

The increasing demand for wireless communication services has necessitated the efficient allocation of spectrum resources. Traditional methods of spectrum management are often static and inefficient, prompting researchers to explore dynamic approaches, such as RL, to optimize spectrum usage. This literature review highlights recent advancements in spectrum resource allocation utilizing RL, drawing from more than ten recent studies.

RL, a branch of ML, provides a framework where agents learn to make decisions by interacting with the environment. RL's applicability to spectrum management has gained significant traction due to its ability to handle complex, dynamic environments. For instance, \cite{bhattacharya2022deep} demonstrated the use of deep Q-networks (DQN) for dynamic spectrum access, showing considerable improvements in spectrum utilization efficiency compared to traditional methods.

Similarly, \cite{kassab2020multi} explored the use of policy gradient methods for spectrum allocation in cognitive radio networks. Their study highlighted that RL could adapt to varying network conditions and effectively allocate spectrum without requiring extensive prior knowledge of the environment. Another notable work by \cite{lei2020deep} introduced an AC framework that balances exploration and exploitation, achieving near-optimal spectrum allocation in real time.

In addition to the fundamental RL approaches, hybrid models combining RL with other techniques have also been explored. For example, the authors in \cite{sun2022joint} proposed a hybrid model integrating RL with game theory to address the issue of interference management in multi-user environments. Their results indicated a significant enhancement in overall network performance. Moreover, \cite{alsulami2022federated} combined RL with FL to enable collaborative spectrum sharing among multiple BS, ensuring data privacy while optimizing resource allocation.

The application of RL in spectrum allocation has also been extended to vehicular networks. In \cite{wang2021intelligent}, the authors investigated the use of RL for spectrum management in vehicular ad-hoc networks (VANETs). Their findings suggest that RL can effectively manage spectrum resources in highly mobile environments, reducing latency and improving communication reliability. Similarly, \cite{davsic2020cooperative} applied multi-agent RL to spectrum allocation in IoT networks, demonstrating improved spectrum efficiency and reduced interference.

Another promising direction is the use of RL in 5G and beyond networks. \cite{tang2020deep} explored the application of RL in 5G heterogeneous networks, focusing on dynamic spectrum sharing between macro and small cells. Their approach resulted in significant improvements in network throughput and user experience. Furthermore, \cite{zhang2020towards} discussed the challenges and potential of using RL for spectrum management in 6G networks, emphasizing the need for scalable and robust RL algorithms to handle the ultra-dense network scenarios anticipated in 6G.

\subsection{Challenges and Open Issues}
Recent research in spectrum resource allocation has focused on addressing several key challenges, including dynamic spectrum allocation, energy efficiency, fairness and QoS, and security.

\textbf{Response time}: 
Minimizing response time is crucial for the performance of wireless networks, especially in applications requiring real-time communication. Recent research has concentrated on optimizing spectrum allocation to reduce latency. For instance, incorporating ML algorithms can predict network conditions to minimize delays \cite{ferdouse2019joint}.

\textbf{Energy efficiency}: Energy consumption remains a critical concern in wireless networks. Researchers have investigated the trade-off between spectrum utilization and energy efficiency in cognitive radio networks. In \cite{song2018spectrum}, the authors propose an optimization framework that minimizes total energy consumption while satisfying the QoS requirements of users, thus ensuring efficient use of energy without compromising performance .

\textbf{Fairness and QoS}: Ensuring fairness among users and meeting their QoS requirements is essential for efficient spectrum utilization. A multi-objective optimization approach has been proposed in \cite{haryadi2017fairness} to address this challenge. This approach considers both fairness and QoS in spectrum allocation, striving to balance the objectives of maximizing throughput and minimizing the delay experienced by users. 

\textbf{Security}: Security is paramount in wireless networks, especially in CR networks where unauthorized access can pose significant threats. In \cite{al2022overview}, the authors have investigated the security challenges unique to CR networks. One study proposes a spectrum allocation scheme that incorporates security considerations, ensuring the integrity and confidentiality of spectrum resources. This approach helps to mitigate risks associated with unauthorized spectrum use and potential malicious attacks.

By addressing these key challenges, recent advancements in spectrum resource allocation aim to create more robust, efficient, and secure networks capable of adapting to the complex and evolving spectrum environment.

%% file: beam_management.tex
\section{Beam Management}\label{sec6}

Beam management is a critical task in modern wireless communication systems, particularly with the advent of 5G and the exploration of 6G technologies. These communication systems increasingly rely on millimeter wave (mmWave) frequencies to meet the growing demand for higher data rates and capacity. However, mmWave signals, while capable of delivering vast amounts of data, are highly susceptible to attenuation, blockages, and rapid fluctuations in the channel quality. To deal with these challenges, efficient beam management, including beam selection (or, equivalently, beam alignment) and beam tracking, is imperative. 

Beam management faces several challenges, including the intrinsic characteristics of mmWave propagation, the dynamic nature of wireless environments, and user mobility. 
Conventional beam management methods typically rely on predetermined models that are based on theoretical understandings of the system dynamics and channel models. These methods often use simplified assumptions about the environment and user behavior. As a result, they have limited adaptability to dynamic and complex environments, and may not perform well when unexpected changes in the network happen. Moreover, conventional beam management methods may struggle to scale in complexity with the growth of the network and the increasing demands on system performance, such as high accuracy and low latency.

Recently, there has been a growing interest in data-driven beam management. 
As they can utilize the data collected from the real world to uncover complex patterns and correlations, they can offer enhanced flexibility and accuracy.
Moreover, they usually excel in real-time performance and scalability as they may handle high-dimensional data and have low computational complexity in predictions.
In this section, we will introduce some data-driven approaches for beam management. Table~\ref{table:BM_review} provide a review on the surveyed literature, which spans four main aspects:
\begin{itemize}
    \item \textbf{Beam selection.} Beam selection, also known as beam alignment, is the process of choosing the best beam direction based on the measurements obtained during beam training. The purpose is to establish the initial communication link with a high quality by aligning the beams optimally between the BS and the UE. Data-driven approaches are used for improving accuracy and efficiency of beam selection~\cite{long2018data, alrabeiah2020deep, ma2020machine, echigo2021deep, ma2021deep, polese2021deepbeam, xu20203d, mashhadi2021federated, alrabeiah2020millimeter}, as well as taking other communication metrics into account~\cite{wu2019fast, zhang2020beam, gupta2020beam}. 

    \item \textbf{Beam tracking.} Beam tracking involves continuously monitoring and adjusting the beam direction to maintain optimal alignment as the UE moves or the environment changes. The purpose is to maintain a high-quality connection over time, adapting to the mobility of the UE and variations in the radio environment. Data-driven approaches are adopted to model and capture the features of mobile UE and environment, and support reliable beam tracking~\cite{xu2021data, burghal2019machine, lim2021deep, shah2023lstm, zhang2020beam, booth2019multi, li2023machine}.
    
    \item \textbf{Codebook design.} Classical beam training process relies on pre-defined codebooks, which consist of fixed beam patterns that are not adaptable to real-time changes in the environment, such as dynamic blockages, user mobility, and varying propagation conditions. To cover all possible directions and scenarios, pre-defined codebooks often include a large number of beams, leading to extensive beam training procedures, since the process of sweeping through numerous beams to find the optimal one can be time-consuming and resource-intensive. Data-driven approaches~\cite{wang2022fast, zhang2021reinforcement, zhang2020intelligent}, especially RL methods~\cite{zhang2021reinforcement, zhang2020intelligent}, are utilized to implement adaptive codebooks with smaller size. 

    \item \textbf{Joint design.} Joint design of beam management and other factors, e.g., power control, interference coordination, and relay selection, is a complex problem, which is usually solved with sophisticated optimization tools based on some knowledge such as channel state. To obviate the need for such knowledge, RL methods are employed~\cite{mismar2019deep, xue2022beam, kim2023joint}. 
\end{itemize}

\begin{table*}[h]
\caption{Review of data-driven approaches for beam management.}
\renewcommand{\arraystretch}{1.5}
\scriptsize
\centering
\scriptsize
\begin{tabular}{|c|c|c|c|c|c|}
\hline
\multicolumn{2}{|c|}{\textbf{Data-driven approach}} & \textbf{Beam selection} & \textbf{Beam tracking} & \textbf{Codebook design} & \textbf{Joint design} \\ \hline
\multicolumn{2}{|c|}{Statistical approach} & -- & -- & \cite{wang2022fast} & -- \\ \hline
\multirow{2}{*}{ML} & SVM & \cite{long2018data} & -- & -- & -- \\ 
~ & GPML & -- & \cite{xu2021data} & -- & -- \\ \hline
\multirow{5}{*}{DL} & DNN & \cite{alrabeiah2020deep} & -- & -- & -- \\  
~ & CNN & \cite{ma2020machine, echigo2021deep, ma2021deep, polese2021deepbeam, xu20203d, mashhadi2021federated} & -- & -- & -- \\ 
~ & LSTM & \cite{echigo2021deep, ma2021deep} & \cite{burghal2019machine, lim2021deep, shah2023lstm, li2023machine} & -- & -- \\  
~ & ResNet & \cite{alrabeiah2020millimeter} & -- & -- & -- \\
~ & Transformer & -- & \cite{li2023machine} & -- & -- \\\hline
\multirow{4}{*}{RL} & DQN & -- & -- & \cite{zhang2020intelligent} & \cite{mismar2019deep} \\ 
~ & DDQN & -- & -- & -- & \cite{xue2022beam} \\ 
~ & DDPG & -- & -- & \cite{zhang2021reinforcement} & \cite{kim2023joint} \\  
~ & Actor-Critic & -- & -- & \cite{zhang2021reinforcement} & -- \\ \hline
\multirow{2}{*}{Online learning} & UCB & \cite{wu2019fast, zhang2020beam, gupta2020beam} & \cite{zhang2020beam} & -- & -- \\ 
~ & TS & \cite{booth2019multi} & \cite{booth2019multi} & -- & -- \\ \hline
\end{tabular}
\label{table:BM_review}
\end{table*}

\subsection{Statistical Approaches for Beam Management}
Statistical approaches introduce a data-centric perspective to beam management. By applying statistical techniques, such as probabilistic modeling, hypothesis testing, etc., it can be more effective to identify the optimal beam.
Wang~et~al.~\cite{wang2022fast} presented a statistical approach for optimizing beam training, which does not require information about the location of incoming UE. Instead, it leverages historical channel data from UEs that the BS has served, which includes angles of arrival and departure, as well as the channel strength, to statistically characterize the spatial distribution of UEs. A density-based spatial clustering algorithm was then used to identify an optimized set of beams that match the spatial distribution of UEs, which can significantly reduce the initial access delay for UEs compared to traditional sequential beam training and omnidirectional approaches. 

\subsection{ML for Beam Management}
Traditional ML, such as SVM, has been applied to predict optimal beam directions based on environmental information and historical data, offering a reduction in complexity and latency. 

In~\cite{long2018data}, the beam selection problem was considered a multi-class classification problem, where substantial samples of mm-wave channel serve as training data. For the sake of low complexity, the SVM algorithm was utilized to address the challenge of analog beam selection for MIMO systems. Specifically, the SVM algorithm was applied to develop a statistical model aimed at optimizing the sum-rate performance. This model enables the selection of the optimal analog beam for each user during real-time communication, significantly simplifying the process.

In unmanned aerial vehicle (UAV) networks, angular domain information (ADI) plays a critical role in optimizing beamforming, which can be inferred from the UAVs' relative positions. However, acquiring UAV positions is time-consuming and incurs significant overhead. Moreover, due to the mobility of UAVs, the position data may become outdated if not updated on time. To tackle this challenge, Xu~et~al.~\cite{xu2021data} utilized GPML to facilitate position prediction. Upon the predicted UAV positions, the UAVs were clustered for coarse ADI acquisition, which would be confined and used for fast beam tracking. Then, GPML was employed again for dynamically selecting beam patterns for various clusters at different times.

However, one of the critical challenges faced by traditional ML algorithms is their limited fitting capability, primarily when relying on relatively simple models. This limitation becomes particularly evident in complex scenarios characteristic of mmWave communications, where the environment's dynamics and the high dimensionality of the data can surpass the capacity of simpler ML models to capture and learn from the underlying patterns effectively.

\subsection{DL for Beam Management}
DL offers several advantages in the context of beam management, which stems from DL's ability to model complex nonlinear factors such as UE's movement and channel variations, adaptively learn from the environment, and effectively handle high-dimensional features of the propagation scenario. 

\subsubsection{High-Resolution Beam Prediction}
DL emerges as a solution for modeling the nonlinear power leakage and multipath interference, facilitating accurate Angle of Arrival (AoA)/Angle of Departure (AoD) estimation. 
By leveraging the inherent spatial characteristics of the communication channels, DL models enable the prediction of the optimal high-resolution beam direction through a preliminary low-resolution beam search, effectively minimizing the training overhead.

\begin{figure}
    \centering
    \begin{subcaptiongroup}
        \centering
        \parbox[b]{\linewidth}{
        \centering
        \includegraphics[width=0.99\linewidth]{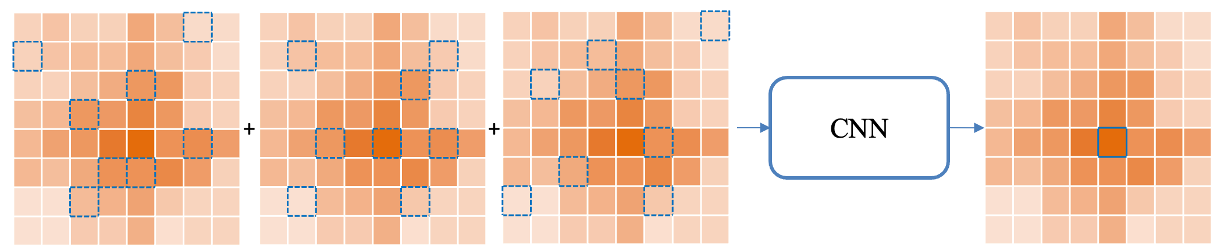}
        \caption{\textrm{Illustration of high-resolution beam selection scheme in~\cite{ma2020machine}.}}\label{fig:subfig1}}%
        \vspace{5mm}
        \parbox[b]{\linewidth}{
        \centering
        \includegraphics[width=0.88\linewidth]{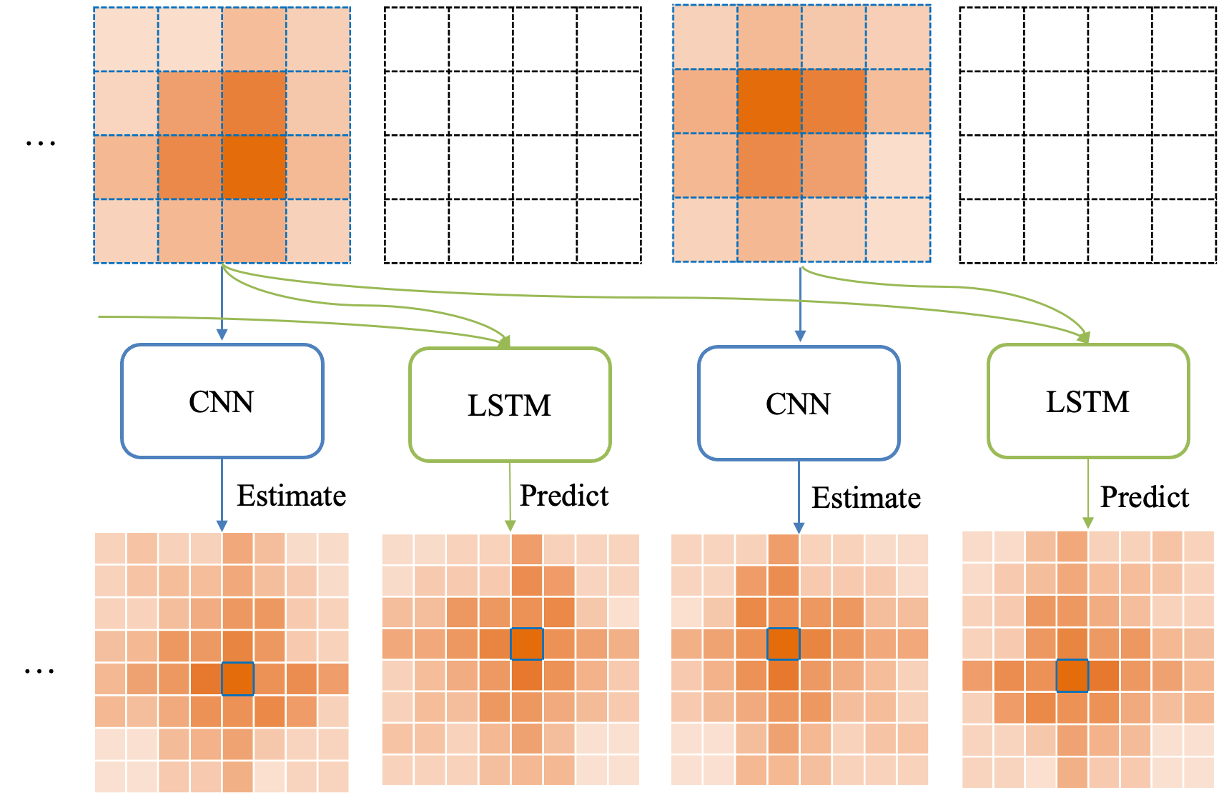}
        \caption{\textrm{Illustration of high-resolution beam selection scheme in~\cite{echigo2021deep}.}}\label{fig:subfig2}}
        \vspace{5mm}
        \parbox[b]{\linewidth}{%
        \centering
        \includegraphics[width=0.85\linewidth]{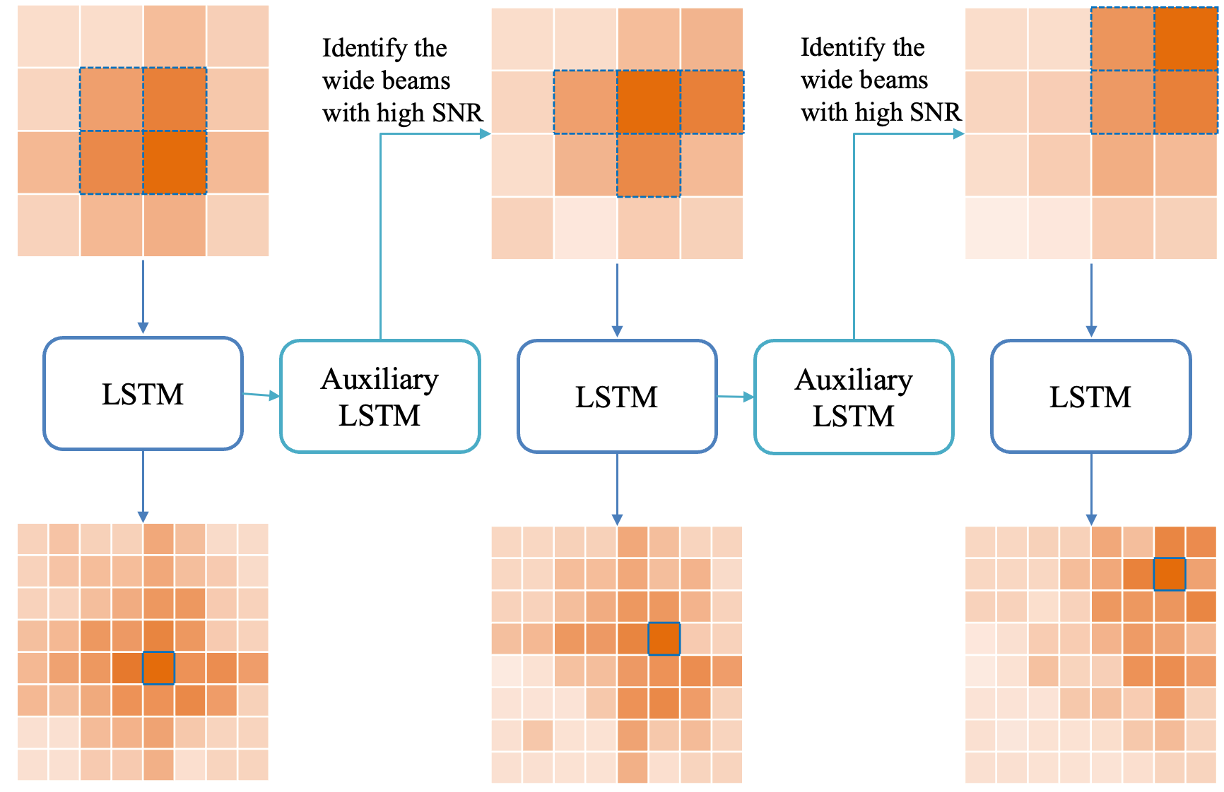}
    \caption{\textrm{Illustration of high-resolution beam selection scheme in~\cite{ma2021deep}.}}\label{fig:subfig3}}
    \end{subcaptiongroup}
    \caption{Illustration of high-resolution beam selection schemes. We consider a codebook with $4\times 4$ wide beams and a codebook with $8\times 8$ narrow beams, which are represented by two square arrays. Each square in the arrays corresponds to the power intensity of one beam in the codebook. A darker colour means a greater intensity. Small squares with blue dashed lines indicate the measured beams, and those with blue solid lines indicate the optimal beams.}
    \label{fig:illustration}
\end{figure}

A pragmatic strategy to reduce beam-training overhead is to select beams with specified angular and utilize their received signals for optimal beam prediction. Given the finite number of beam pairs, this task can be transformed into a classification problem.
Ma~et~al.~\cite{ma2020machine} adopted a CNN classifier, followed by a softmax layer that assigns probabilities to each beam pair. Subsequently, the pair with the highest probability can be chosen. During the training phase, all candidate beams were swept to label the optimal beam. To gather prediction inputs, only a subset of all candidate beams in the codebook were swept, and the received signals were used as the network input. Without sweeping all beams in the codebook in every iteration, the beam training overhead can be reduced. Upon completion of the offline training, the model was applied online to predict the beam distribution vector, which enables the simultaneous alignment of beams for all users, guided by the dominant entries within the beam distribution vector. 
In~\cite{echigo2021deep}, the high-resolution beam prediction was regarded as a super-resolution image recovery problem. The proposed method leveraged wide-beam measurements to infer the properties of narrow beams. It utilizes a CNN to capture the spatial correlation among beam qualities so that the full signal space can be covered with low overhead. 
Additionally, an LSTM network was employed in the prediction to take the temporal relationship of the beam qualities into account. The proposed LSTM architecture can utilize past beam measurements and maintain only useful information, and hence can further reduce overhead. However, when the AoD of the dominant path is not located in the main lobe of any sampled beam, the performance may be significantly degraded due to low signal-to-noise ratio (SNR).
Therefore, Ma~et~al.~\cite{ma2021deep} proposed two criteria to select the wide beams with high SNR for the subsequent training based on the received signals from previous beam training, which helps further reduce the overhead of wide beam training. An LSTM network was implemented to predict the current optimal narrow beam, and an auxiliary LSTM module LSTM module integrating these proposed criteria was adopted to identify the wide beams with high SNR more precisely. 
The above high-resolution beam selection schemes are illustrated in Fig.~\ref{fig:illustration}.

\subsubsection{Side-Information-Aided Beam Selection}
DL is employed to assimilate high-dimensional environmental features derived from side information concerning the environment. 
To reduce hardware costs, mmWave antennas may be employed by low-frequency BSs. In such co-located setups, both the low-frequency and mmWave links operate within similar propagation environments, allowing the mmWave channel to share comparable AoA AoD characteristics with its low-frequency counterpart. Consequently, DL can be leveraged to uncover the intricate relationships between low-frequency and mmWave channels, taking advantage of their common environmental traits. 
In~\cite{alrabeiah2020deep}, a fully connected DNN was utilized to predict optimal mmWave beams and blockage status from sub-6 GHz channel information. It was shown that sufficiently large neural networks can accurately predict the optimal mmWave beams and blockage, achieving an arbitrarily high success probability. 
Recognizing that different beam patterns introduce distinct impairments in the waveform, 
in~\cite{polese2021deepbeam}, a CNN-based beam management solution called DeepBeam was introduced. DeepBeam utilizes the feature extraction capabilities of a CNN to infer the AoA and beam by simply monitoring the ongoing transmissions within the network. 
Specifically, a device with DeepBeam can collect the in-phase and quadrature (I/Q) data from the physical layer, which is used as the input of the CNN to infer the best beam pair.

Visual imagery may also serve as a valuable source of environmental information to facilitate beam management. Alrabeiah~et~al.~\cite{alrabeiah2020millimeter} introduced a novel approach that leverages images captured by cameras installed at BSs to predict the optimal beam, eliminating the requirement of channel measurements or beam-training procedures. To achieve this, the ResNet architecture was employed to learn relevant features from the images. 
Furthermore, Xu~et~al.~\cite{xu20203d} explored the potential of a 3D scene-based beam selection framework, which begins with the construction of a 3D scene point cloud that covers the cellular network's service area. Following this, the BS integrated the UE's geographical location with the point cloud to synthesize panoramic scene data. A 3D-CNN was then utilized to predict the optimal beam based on this panoramic information. 

\subsubsection{Beam Tracking}
Beam tracking involves tracking the optimal beamforming direction as the environment changes. This task requires processing sequential data, such as time-varying channel conditions, where historical information and temporal dependencies play a crucial role.  
LSTMs are specifically designed to handle sequential data by capturing short-term (e.g., multipath fading) and long-term (e.g., shadowing and blocking) dependencies, which makes LSTMs well-suited for beam tracking tasks.

 The overhead of beam management is significantly increased by the mobility of UE. As users move, the optimal beam direction for maintaining high-quality communication links continually shifts, necessitating frequent beam adjustments, i.e., beam tracking.
 Nonlinearity is a characteristic feature of user mobility patterns in wireless networks, which exacerbates the challenges associated with time-varying channels. In~\cite{burghal2019machine}, LSTM was used to end-to-end estimate the AoA for beam tracking, where the input includes the noisy channel observation and the previous estimate of the AoA, and the output is an estimate of the current AoA. To incorporate both model-based and data-driven approaches, Lim~et~al.~\cite{lim2021deep} only replaced the model-based prediction step in the Bayesian filtering framework with LSTM. 
 Due to the high susceptibility of mmWave signals to blockages such as human bodies and buildings, dense millimeter-wave cell deployments are necessary to provide macro-diversity. As a result, UEs need to track multiple beams from multiple cells. In~\cite{shah2023lstm}, multi-cell multi-beam tracking under power/overhead constraints was investigated. This problem can be formulated as a Partially Observed MDP (PO-MDP) and is usually solved by RL methods, which require learning a Q-function. However, due to the deployment of dense cells and numerous antennas, the action and state space of such problems can be enormous. Therefore, an LSTM-based supervised learning method was proposed to approximate the action sequence, which avoids the high computation complexity induced by learning the Q-function. 

 In recent years, Transformer, originally developed for natural language processing tasks, has also shown promising results when applied to beam tracking problems.  
 In~\cite{li2023machine}, Transformer was used to predict reference signal received power (RSRP), based on which beam tracking can be executed without frequent measurements. As Transformer uses a self-attention mechanism to model the dependencies between different parts of the input, without relying on the sequential order of the data as required by LSTM, it can capture long-range dependencies more effectively than LSTM, which can be particularly useful for complex beam behavior.

\begin{table*}[ht] 
	\caption{Review of DL approaches for beam tracking.}
 \centering
	\label{tab: RL_LA}
 \scriptsize
\resizebox{\textwidth}{!}{
	\begin{tabular}[c c c c]{|m{0.15\textwidth}<{\centering}|m{0.15\textwidth}<{\centering}|m{0.15\textwidth}<{\centering} |m{0.15\textwidth}<{\centering}| m{0.3\textwidth}<{\centering}|}
 
\hline
\textbf{Reference} & \textbf{DL method} & \textbf{Input} & \textbf{Output} & \textbf{Key features} \\ \hline
{Burghal~et~al.~\cite{burghal2019machine}} & LSTM & Noisy channel observation (signal) \& previous estimate of AoA & Estimate of current AoA & Downlink single user;
End-to-end training\\ \hline
Lim~et~al.~\cite{lim2021deep} & LSTM & Sequence of previous estimates of AoA and AoD \& IMU sensor signals
  & Distribution of current AoA \& AoD & Downlink single user;
Replace the model-based prediction step in the Bayesian filtering framework with LSTM
 \\ \hline
 Shah and Rangan~\cite{shah2023lstm} & LSTM  & Channel quality (SNR) of $K$ links
 & 1. $K$ links to be observed;
2. Maximum channel quality
 & Multi-cell single user;
Power/overhead constraints;
POMDP modeling;
Use LSTM-based supervised learning method to approximate the action sequence;
Maximize observed SNR
\\ \hline
Li~et~al.\cite{li2023machine} & Transformer \& LSTM & Reference signal received power (RSRP)
 & \textit{Transformer/LSTM}: predicted future RSRP;
\textit{LSTM}: to change beam or not
 & Track the best beam based on the predicted RSRP/Predict whether the future best beam or beam pair will be different from the one identified at the latest measurement occasion
 \\ \hline
\end{tabular}
\label{table:BM_beam_tracking}
}
\end{table*}

\subsection{RL for Beam Management}
Wireless communication environments are highly dynamic, with frequent changes in user positions, channel conditions, and network loads. RL stands out as a powerful tool for beam management by continuously learning from the interaction with environments, allowing communication systems to dynamically adjust beamforming strategies to maintain optimal communication links. 
DRL combines DL techniques with RL. Typically, deep RL models use DNNs as function approximators to represent the value function or policy. 

\begin{figure}[h]
    \centering
    \includegraphics[width=0.9\linewidth]{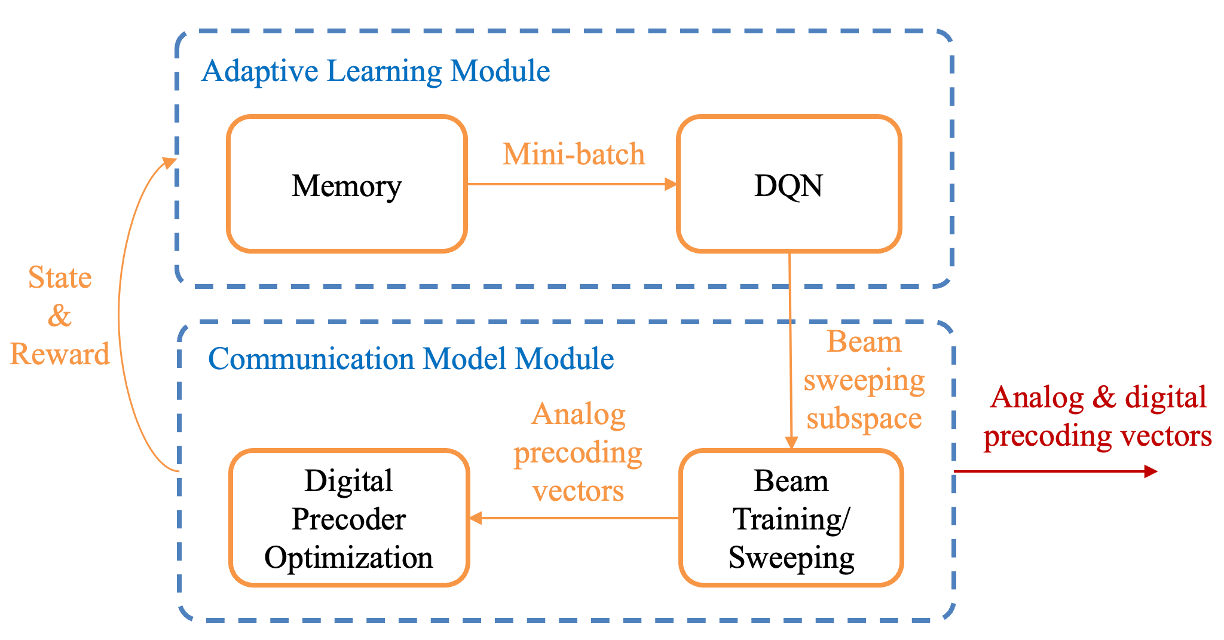}
    \caption{Illustration of beam management scheme in~\cite{zhang2020intelligent}.}
    \label{fig:drl1}
\end{figure}

Classical beamforming usually relies on pre-defined beam codebooks, which consist of numerous narrow beams without adaptation to the surrounding environments. The large codebook sizes lead to high beam training overheads. In~\cite{zhang2021reinforcement}, an adaptive codebook designing method was proposed, which leverages a DRL framework and does not require any knowledge about the channel, RF hardware, or user positions. Then, a Wolpertinger-variant architecture was utilized to reduce the action space and hence learning time. The reward depends on if the quality of the beam exceeds an adaptive threshold.
Considering multi-user scenarios, Zhang~et~al.~\cite{zhang2020intelligent} incorporated both the optimization technique, which utilizes the domain knowledge of wireless communications, and the DRL technique, which has the adaptability to dynamic environments (see Fig.~\ref{fig:drl1}). 
Specifically, the digital precoders and the analog precoders were derived by solving an optimization problem and beam sweeping, respectively. The role of DQN was to provide a smaller beam-sweeping subspace. The reward was designed to be related to the optimal value of the optimization problem, and the duration of learning, precoding, and transmitting.
To enhance communication performance, it is crucial to jointly consider beam management with other factors, such as power control, interference coordination, relay selection, etc. 
Mismar~et~al.~\cite{mismar2019deep} formulated a non-convex optimization problem to jointly optimize beamforming, power control, and interference coordination in multi-access networks with multiple BSs. To solve this tricky problem, they represented it with an MDP and utilized DQN to obtain actions, which consist of the beamforming vector and the transmit power of BSs. The reward was related to the effective SINR of the downlink transmission and its distance from the target SINR. 
Kim~et~al.~\cite{kim2023joint} investigated the joint beam management and relay selection problem in mobile mmWave networks such as vehicle networks, aiming to select relays with minimal beam training overhead by utilizing the deep deterministic policy
gradient (DDPG) algorithm. The realized spectral efficiency was selected as the reward.

\subsection{Online Learning for Beam Management}



Beam management can also be considered an MAB problem, where the selection of different beams can be seen as the choice of different arms. The performance of each beam (or arm) is uncertain and may vary due to changing channel conditions, user mobility, and interference.

In~\cite{wu2019fast, zhang2020beam}, the UCB algorithm was utilized to deal with the beam selection and tracking problems, which are formulated as stochastic bandit problems. 
In~\cite{wu2019fast}, the authors proposed a hierarchical beam selection scheme that leveraged beam correlation and prior knowledge of channel fluctuations to maximize the cumulative received signal strength over a given time period. By exploiting these additional contextual factors, their UCB-based approach was able to more efficiently explore the beam selection space.
In~\cite{zhang2020beam}, the focus shifted to maximizing the expected cumulative effective achievable rate. Rather than defining individual beams or codewords as separate arms in the bandit problem, they defined the arms based on the beam index difference or offset between the two optimal beams in adjacent time slots. This alternative formulation allowed their UCB-based solution to better capture the inherent temporal correlation present in the optimal beam selections. 

However, they considered stationary bandit models which are only suitable for time-invariant channels~\cite{wu2019fast, zhang2020beam}. To tackle non-stationarity induced by time-varying channels, Gupta~et~al.~\cite{gupta2020beam} introduced sliding windows and discounting into the UCB algorithm, with the objective of minimizing outage probability. 
Other than the UCB framework, in~\cite{booth2019multi}, Kalman filter and Linear Thompson sampling were integrated to track time-varying mmWave channels. 

\begin{table*}[h]
\caption{Review of online learning approaches for beam management.}
\renewcommand{\arraystretch}{1.5}
\resizebox{\linewidth}{!}{ 
\centering
 \scriptsize
\begin{tabular}{|m{0.1\textwidth}<{\centering}|m{0.1\textwidth}<{\centering}|m{0.15\textwidth}<{\centering}|m{0.1\textwidth}<{\centering} |m{0.15\textwidth}<{\centering}| m{0.3\textwidth}<{\centering}|}
\hline
\textbf{Reference} & \textbf{Algorithm} & \textbf{Task} & \textbf{Non-stationarity} & \textbf{Output} & \textbf{Key features} \\ \hline
Wu~et~al.~\cite{wu2019fast} & UCB & Beam selection & $\times$ & Optimal beam & Leverage beam correlation and prior knowledge of channel fluctuations;
Maximize the cumulative received signal strength over a specific time period
 \\ \hline
Zhang~et~al.~\cite{zhang2020beam} & UCB & Beam selection \& tracking & $\times$ & Optimal beam & Maximize the expected cumulative effective achievable rate
 \\ \hline
Gupta~et~al.~\cite{gupta2020beam} & UCB  & Beam selection
 & $\surd$ & Optimal beam
 & Introduce sliding windows and discounting into the UCB algorithm; Minimize outage probability
\\ \hline
Booth~et~al.~\cite{booth2019multi} & Thompson sampling & Beam selection \& tracking
 & $\surd$ & Channel estimates
 & Integrate Kalman filter and Linear Thompson sampling (LTS) to track time-varying mmWave channels
 \\ \hline
\end{tabular}
\label{table:BM_OL}
}
\end{table*}

\subsection{Distributed Learning for Beam Management}

In dynamic wireless environments, timely decision-making is crucial for effective beam management. Centralized approaches may suffer from high latency due to the need for data transmission and processing between nodes and a central entity. 
Moreover, in traditional centralized approaches, raw data from multiple nodes is collected and processed centrally, which raises privacy concerns as sensitive information from different nodes may be exposed. 
Distributed learning methods allow local learning and model updates to occur at individual nodes without the need for sharing raw data, which reduces communication overhead and latency, and preserves the privacy of sensitive user data while still benefiting from collaborative learning.

Xue~et~al.~\cite{xue2022beam} tackled the challenge of beam management in heterogeneous ultra-dense mmWave communication systems. The authors modeled this problem as an MDP, which had a particularly large state-action space. To address the complexity introduced by this large space, as well as concerns around data privacy, they utilized a federated RL framework.
Their approach worked as follows: At the start of each training round, the macro BS would transmit the current state of the global beam management model to the individual mmWave BSs, each of which would then independently use a double deep Q-network (DDQN) to train a local model. These locally-trained updates were then sent back to the macro BS, which incorporated them into the global model using the FederatedAveraging (FedAvg) approach~\cite{mcmahan2017communication}.
In~\cite{mashhadi2021federated}, the FedAvg approach was also leveraged to assist with beam selection in vehicle-to-infrastructure (V2I) mmWave communication systems. The authors focused on LiDAR-aided beam selection, where connected vehicles collaboratively trained a shared neural network model using their local location data and LiDAR measurements.

\subsection{Challenges and Open Issues}
While data-driven approaches offer significant advantages for beam management, they also present several challenges and open issues that need to be addressed.
\begin{itemize}
    \item \textbf{Dataset quality and availability.} Inaccurate data and incomplete datasets can lead to poor training performance and hence unreliable beam management. 
    In beam management, training is usually collected using search-based methods. 
    The substantial overhead involved in collecting relevant data presents a practical challenge. 
    To address the challenge of data collection, semi-supervised learning and few-shot learning should be considered. 
    

    \item \textbf{Energy efficiency.} The computational demands of data-driven approaches can lead to high power consumption, which is a concern for energy efficiency. It is necessary to balance the need for powerful processing capabilities with the goal of maintaining sustainable and energy-efficient operations.

    \item \textbf{Multiple users with high mobility.} Beam management in high-mobility scenarios is particularly challenging due to the rapid changes in user positions and the environment. Managing beams for multiple users simultaneously while ensuring optimal performance and minimal interference is complex and resource-intensive.
    Therefore, it's a worthwhile study to manage interference in high-density user environments, ensuring each user receives signals with high quality.

    \item \textbf{Collaborative edge AI.} Collaborative edge AI is anticipated to be a crucial component in 6G networks. Leveraging the resources located at multiple users, collaborative edge AI enables adaptive, low-latency beam management strategies. 
    Moreover, collaborative paradigms, such as federated learning, not only optimize resource allocation but also address privacy and security concerns. Continued research and industry collaboration will be vital for harnessing its full potential. 
    
    \item \textbf{Multi-modal integration of side information.} Although various side information has been utilized for beam management~\cite{alrabeiah2020deep, polese2021deepbeam, alrabeiah2020millimeter, xu20203d}, each of the studies only uses a single data source.
    By leveraging diverse data sources such as cameras, LiDAR, and IoT sensors, multi-modal integration provides a richer, more accurate understanding of the environment, leading to robust and context-aware beam management. 
\end{itemize}

%% file: power_control.tex
\section{Power Control}
\label{sec7}
\subsection{Introduction}
Power control is a critical technique in wireless communication systems, especially in the context of dense and heterogeneous network deployments such as 5G and beyond. Advanced power control mechanisms are needed to dynamically adapt to changing network conditions and user demands, ensuring that interference is minimized and resources are utilized efficiently. With the advent of ML and artificial intelligence, data-driven methods have emerged as promising alternatives to traditional power control strategies. These methods can leverage large volumes of data to learn complex patterns and make real-time decisions, offering improved adaptability and performance in dynamic and heterogeneous network environments. By incorporating advanced algorithms and neural network architectures, these approaches can effectively address the challenges of power control in modern wireless communication systems.

\subsubsection{Objectives and Constraints}
The main objectives of power control include:
\begin{itemize}
\item \textbf{Maximize Network Throughput:} Enhance the sum-rate of data transmission across the network.
\item \textbf{Maximize Energy Efficiency:} Optimize power allocation to maximize data transmission while minimizing energy consumption.
\item \textbf{Minimize Interference:} Reduce the impact on adjacent cells and networks to promote harmonious operation.
\end{itemize}
These objectives are subject to the following constraints:
\begin{itemize}
\item \textbf{Power Budget:} Adhere to the maximum transmission power limit for each UE.
\item \textbf{QoS:} Meet the minimum SINR requirements to uphold the predefined service quality.
\end{itemize}
Additionally, the optimization problem must address the non-convexity inherent in the system and ensure real-time implementation feasibility while seeking low-complexity and scalable solutions.

\subsubsection{Limitations and Challenges}
\label{ssec:limitations}
The non-convexity of the power control optimization problem is a fundamental challenge. These optimization problems are inherently non-convex due to the complex coupling between the transmit powers and the resulting SINRs at the receivers. The high computational complexity of these traditional methods becomes a bottleneck.

Traditional power control algorithms, such as the WMMSE and fractional programming (FP) methods, rely on instantaneous global channel CSI availability to optimize the transmit powers of the Access Points (APs). However, in dynamic wireless environments characterized by rapid channel condition fluctuations, the processing time required for these centralized optimization-based algorithms often exceeds the channels' coherence time, rendering their solutions obsolete. Therefore, the necessity for instantaneous global CSI poses significant practical challenges in large-scale and decentralized communication networks, where acquiring and maintaining up-to-date CSI can be prohibitively complex.

In dynamic environments where the network topology and channel conditions change rapidly, traditional methods may fail to adapt quickly enough to maintain optimal performance. As the number of UEs increases, the complexity of traditional algorithms grows, leading to scalability issues in large-scale networks. Additionally, ensuring energy efficiency alongside performance optimization poses another significant challenge, especially in resource-constrained environments.

\subsection{DL for Power Control}

\begin{table}[t]
    \caption{Comparison of DL approaches for power control}
    \renewcommand{\arraystretch}{1.5}
    \resizebox{\linewidth}{!}{ 
    \centering
    \scriptsize
    \begin{tabular}{|m{0.12\textwidth}<{\centering}|m{0.12\textwidth}<{\centering}|m{0.12\textwidth}<{\centering} |m{0.12\textwidth}<{\centering}| m{0.12\textwidth}<{\centering}|m{0.12\textwidth}<{\centering}|m{0.12\textwidth}<{\centering}|}
    \hline
        Method/Objective & Maximize Network Throughput & Maximize Energy Efficiency & Minimize Interference \\ \hline
        DNN & - PowerNet (Chien et al.~\cite{Van2020}) & ~ & - CDNN, DDNN, DDNN-SI (Zaher et al.~\cite{Zaher2023}) \\ 
        ~ & - Unsupervised DNN (Rajapaksha et al.~\cite{rajapaksha2021deep}) & ~ & ~ \\  \hline
        GNN & - HGNN (Li et al.~\cite{li2024heterogeneous}) & - MMSE-GNN (Peng et al.~\cite{peng2023learning}) & - HGNN (Li et al.~\cite{li2024heterogeneous}) \\
        ~ & -CF-HGNN (Li et al.~\cite{li2024heterogeneous}) & ~ & - CF-HGNN (Li et al.~\cite{li2024heterogeneous}) \\ 
        ~ & -HetGNN (Guo et al.~\cite{guo2022}) & ~ & ~ \\ \hline
        RNN & -DRNN (Guo et al.~\cite{guo2023deep}) & - DRNN (Guo et al.~\cite{guo2022,guo2023deep}) & - IFE-RNN (Shi et al.~\cite{shi2021distributed}) \\ 
        ~ & -IFE-RNN (Shi et al.~\cite{shi2021distributed}) & ~ & ~ \\ \hline
    \end{tabular} 
\label{DL_power}
}
\end{table}

DL techniques have been utilized to model the complex non-linear relationships between transmission power and network performance metrics. Incorporating data rate models within DL frameworks has emerged as a pivotal strategy for enhancing the efficiency of power allocation in wireless communication systems. Among these techniques, DNNs can operate in both centralized and decentralized environments and handle both supervised and unsupervised learning tasks. The main advantage of DNNs is their relatively low inference complexity, making them suitable for real-time applications post-training. However, they have high training complexity, which can be a drawback in some scenarios. GNNs, with moderate complexity, leverage graph structures to effectively model data relationships and offer good scalability, making them ideal for network topology modeling. The key benefit of GNNs is their ability to capture the complex interdependencies within the network structure, leading to improved performance in power allocation and network optimization tasks. However, their training process can be more computationally intensive compared to simpler models. RNNs are lightweight and low in complexity, excelling in capturing temporal dynamics, which is crucial for sequential data processing. This makes RNNs well-suited for power allocation in dynamic environments where temporal patterns play a significant role. The main advantage of RNNs is their efficiency and suitability for real-time applications. Table \ref{DL_power} provides a comprehensive comparison of all related papers, highlighting the application of these DL techniques, and the details are as follows:

Following their previous work~\cite{guo2022}, Guo and Yang expanded their research to include heterogeneous GNNs (HGNNs) for learning power allocation policies in multi-cell and multi-user systems~\cite{guo2023deep}. The proposed data rate-based neural network (DRNN) features an iterative structure with multiple update layers, leveraging data rate models and permutation equivariance (PE) properties. This approach significantly reduces training samples and time while achieving the desired system performance, as demonstrated by simulation results. The key advantage of the DRNN is its ability to efficiently incorporate data rate models and leverage graph structures, leading to improved power allocation decisions.

In a related advancement, Rajapaksha et al.~\cite{rajapaksha2021deep} proposed a DL-based power control algorithm for cell-free massive MIMO networks. This unsupervised learning approach simplifies the training process and offers a flexible model that can adapt to changing environments, resulting in a significant reduction in implementation time and comparable performance to optimization-based algorithms. The main benefit of this approach is its adaptability to dynamic network conditions and the reduced complexity of the training process.

Continuing the theme of optimizing power control in Massive MIMO systems, Chien et al.~\cite{Van2020} developed PowerNet, a CNN that predicts optimal pilot and data powers from large-scale fading coefficients. PowerNet's innovation lies in its ability to handle dynamically varying active users per cell, offering a substantial improvement over prior methods with a runtime of only 0.03 ms on a GPU. The primary advantage of PowerNet is its computational efficiency, making it suitable for real-time power control in large-scale Massive MIMO systems.

Zaher et al.~\cite{Zaher2023} further explored downlink power allocation in cell-free Massive MIMO systems, addressing both sum SE maximization and proportional fairness. They developed DNN solutions that rely on local large-scale fading information, including a fully distributed DNN (DDNN), a distributed DNN with side information (DDNN-SI), and a clustered DNN (CDNN). These models contribute to a significant reduction in computational complexity and signaling overhead, making them attractive for practical implementation in cell-free Massive MIMO networks.

In the domain of D2D networks, Shi et al.~\cite{shi2021distributed} proposed the IFE-RNN algorithm for power control. This algorithm uses interference feature extractors (IFEs) and an RNN to formulate power allocation as a Markov decision problem, enabling the prediction of real-time interference and informed power decisions with promising performance and computational efficiency. The key benefit of the IFE-RNN is its ability to capture the temporal dynamics of interference in D2D networks, leading to effective power control decisions.

Building on these contributions, Peng et al.~\cite{peng2023learning} introduced an MMSE-GNN for uplink power control in multi-antenna systems. This model integrates the MMSE receiver with graph neural networks to learn optimal power control policies. The MMSE-GNN enhances spectral efficiency and reduces training complexity, showing improved learning performance and generalization ability, particularly when the BS has only partial channel information. The main advantage of the MMSE-GNN is its ability to effectively leverage both the MMSE receiver and the graph structure to optimize power control, resulting in enhanced performance and reduced training complexity.

Lastly, Li et al.~\cite{li2024heterogeneous} introduced a HGNN, known as CF-HGNN, for power allocation in multicarrier-division duplex (MDD) cell-free massive MIMO (CF-mMIMO) systems. The CF-HGNN uses meta-path mechanisms for efficient interference management and adaptive node embedding for scalability across different network sizes, surpassing traditional methods in spectral efficiency and computational complexity. The key benefits of the CF-HGNN are its ability to effectively model the heterogeneous network structure and its scalability to handle varying network sizes, leading to improved power allocation decisions and reduced computational requirements.

\subsection{RL for Power Control}
\begin{table}[t]
    \caption{Comparison of RL approaches for power control}
    \renewcommand{\arraystretch}{1.5}
    \resizebox{\linewidth}{!}{ 
    \centering
    \scriptsize
    \begin{tabular}{|m{0.12\textwidth}<{\centering}|m{0.12\textwidth}<{\centering}|m{0.12\textwidth}<{\centering} |m{0.12\textwidth}<{\centering}| m{0.12\textwidth}<{\centering}|m{0.12\textwidth}<{\centering}|m{0.12\textwidth}<{\centering}|}
    \hline
        Method/Objective & Maximize Network Throughput & Maximize Energy Efficiency & Minimize Interference \\ \hline
        Q-Learning & Q-DPA (Amiri et al. ~\cite{amiri2019reinforcement}) & -Q-DPA (Amiri et al. ~\cite{amiri2019reinforcement}) & ~ \\ 
        ~ &  -DRL(Meng et al.~\cite{meng2020power}) & -DRL (Tan et al.~\cite{Tan2021}) & ~ \\ 
        ~ &  ~ &  -DQN(Su et al.~\cite{su2021})  & ~ \\  \hline
        DDPG & - DDPG (Luo et al.~\cite{luo2022downlink}) & ~ & - DDPG (Wang et al.~\cite{wang2021joint})  \\
       ~ & - DDPG (Meng et al.~\cite{meng2020power}) & ~ & ~  \\ \hline
        MADRL & - MASC (Zhang et al.~\cite{zhang2021deep}) & - MADRL (Xu et al.~\cite{xu2023distributed}) & - MADRL (Xu et al.~\cite{xu2023distributed}) \\ 
        ~ & - MADRL (Xu et al.~\cite{xu2023distributed}) & - MADRL (Sharma et al.~\cite{sha2019}) & MADRL (Li et al.~\cite{li2022}) \\ \hline
    \end{tabular}
    }
\label{RL_power}
\end{table}

RL has become a promising paradigm within ML, particularly for applications in wireless communications. Through repeated interactions with the environment and trial-and-error learning, RL can autonomously address complex decision-making problems, making it highly suitable for overcoming various challenges in communication networks. Specifically, DRL has shown significant potential for power control in wireless networks. DRL algorithms learn optimal policies directly from environmental interactions without the need for instantaneous global CSI, which is often difficult to acquire. DRL-based power control algorithms offer several key advantages: managing large state-action spaces using DNNs, supporting decentralized execution with both centralized and distributed training, achieving superior throughput and energy efficiency compared to traditional methods, and reducing computational complexity by leveraging learning instead of optimization.

Among the various RL techniques, Q-learning is noted for its moderate training complexity and relatively simple implementation, making it a practical choice for many applications. The main benefit of Q-learning is its simplicity and ease of implementation, but it may struggle to scale to large and complex state-action spaces. DDPG also has moderate training complexity but is highly adaptable to complex dynamic environments, which is crucial for real-world applications. DDPG's ability to handle continuous action spaces and its flexibility in adapting to changing conditions are its key strengths. Multi-agent deep reinforcement learning (MADRL) offers scalability, decentralized decision-making, and robustness to uncertainties, making it well-suited for large-scale and dynamic network environments. The main advantages of MADRL are its ability to handle multi-agent scenarios, its decentralized decision-making capabilities, and its resilience to uncertainties in the network. Table \ref{RL_power} comprehensively compares all related papers and the details are as follows:

Amiri et al.~\cite{amiri2019reinforcement} presented a distributed framework based on multi-agent MDPs, introducing a Q-learning algorithm for dynamic power allocation in dense two-tier heterogeneous networks (HetNets). The key benefit of this Q-learning-based distributed power allocation algorithm (Q-DPA) is its ability to dynamically adapt transmit power as new small cells are introduced into the network, ensuring that the macrocell users maintain their required QoS. However, the simplicity of Q-Learning may limit its scalability to larger and more complex network scenarios.

Building upon this foundation, Zhang and Liang~\cite{zhang2021deep} introduced a DRL approach that employs a centralized training and distributed execution model. This method leverages a multiple-actor shared-critic (MASC) structure, which not only optimizes the sum-rate performance but also ensures robustness against the variability of network conditions. The key advantage of this approach is its ability to balance sum-rate optimization and robustness to network dynamics, making it suitable for real-world deployments.

In a related study, Wang et al.~\cite{wang2021joint} introduced a DDPG-based algorithm for joint interference alignment and power control in dense HetNets. The scheme optimizes spectrum efficiency by aligning intra-cell interference and controlling inter-cell interference through centralized power management at the macro BS. The primary benefit of this DDPG-based approach is its ability to handle continuous action spaces and adapt to complex dynamic environments, leading to improved spectral efficiency in dense HetNet scenarios.

Furthermore, Xu et al.~\cite{xu2023distributed} proposed a MADRL scheme with a penalty-based Q learning (PQL) algorithm. This scheme fosters cooperation among agents by incorporating regularization terms in the loss function, leading to a more efficient and coordinated power control strategy that adapts to the dynamic nature of user locations. The key advantages of this MADRL approach are its scalability, decentralized decision-making, and robustness to network uncertainties, making it well-suited for large-scale and dynamic network environments.

In D2D networks~\cite{Tan2021}, a multi-agent DRL algorithm using AC methods was used for joint channel selection and power control. Each D2D pair acts independently, making decentralized decisions based on local observations. This DRL approach significantly improves the sum rate and energy efficiency compared to traditional methods, thanks to its ability to adapt to the dynamic nature of wireless communication systems.

Concurrent with the push for utility maximization, minimizing interference and enhancing fairness have become paramount. Meng et al.~\cite{meng2020power} explored dynamic downlink power control within multi-user cellular networks, comparing the efficacy of various DRL algorithms. Their work underscores the importance of balancing sum-rate maximization with fairness considerations, aiming for a robust and equitable network. Luo et al.~\cite{luo2022downlink} addressed the downlink power control challenge in cell-free massive MIMO systems with a model-free DDPG algorithm. This method not only achieves a commendable performance-complexity trade-off but also ensures equitable user rate distribution, thereby enhancing network fairness.

Recent advancements of RL applied to IoT networks have demonstrated significant potential in optimizing power control and communication efficiency. Li et al.~\cite{li2022} explored the use of MADRL to optimize flight control and data aggregation in Internet-of-Drones (IoD) networks, leveraging a LSTM-based RL model. The MADRL framework ensures that drones can adapt to dynamic network conditions and optimize their operations collectively, leading to reduced data loss and improved communication efficiency. Su et al.~\cite{su2021} focused on optimizing cooperative relaying and power control in Internet of Underwater Things (IoUT) networks, employing an RL learning approach that allows for dynamic adjustment of transmission power and relay selection, thereby improving overall network efficiency and reliability. Sharma et al.~\cite{sha2019} addressed the problem of distributed power control in large-scale energy harvesting networks, proposing a multi-agent MADRL framework that leverages the capabilities of DQN to manage power control across multiple agents, enabling efficient energy usage and improved network performance.

\subsection{Challenges and Open Issues}
This section has provided an overview of power control advancements enabled by DL and RL learning techniques. DL and DRL have emerged as promising solutions, offering improved adaptability and scalability compared to classical optimization-based approaches.

Looking ahead, several key research directions offer opportunities for further advancement:
\begin{itemize}
    \item \textbf{Advanced ML models for non-convex optimization}: Develop sophisticated ML models that can effectively tackle non-convex optimization problems inherent in power control, offering near-optimal solutions with reduced computational complexity.
    \item \textbf{Proactive power control strategies}: Design proactive power control strategies leveraging predictive analytics to adapt to the dynamic nature of wireless networks, ensuring consistent QoS and minimizing interference.
    \item \textbf{Decentralized and distributed power control algorithms}: Investigate decentralized and distributed algorithms for power control that can operate efficiently over large-scale networks with limited central coordination.
    \item \textbf{Robustness against uncertainty and imperfections}: Enhance the robustness of power control algorithms against model uncertainty, channel estimation errors, and hardware imperfections.
\end{itemize}

Addressing these future research directions can drive further evolution of power control, enabling more efficient, adaptive, and intelligent resource management in next-generation wireless networks.

%% file: codesign.tex
\section{Control-Communication Co-design}\label{sec8}
In this section, we review data-driven methods for control and communication co-design problems. We summarize the contributions and limitations of various studies, compare different works, and propose future research directions.

\subsection{Background}
Recent innovations in wireless networking, sensing, computing, and control are revolutionizing the interaction between control systems and informational processes, including CPSs and IoT. Wireless networked control systems (WNCS) are considered a crucial solution for the scalable and cost-effective implementation of widespread automatic control systems. 

\begin{figure}[h]
    \centering
    \captionsetup{justification=centering}
    \includegraphics[scale=0.33]{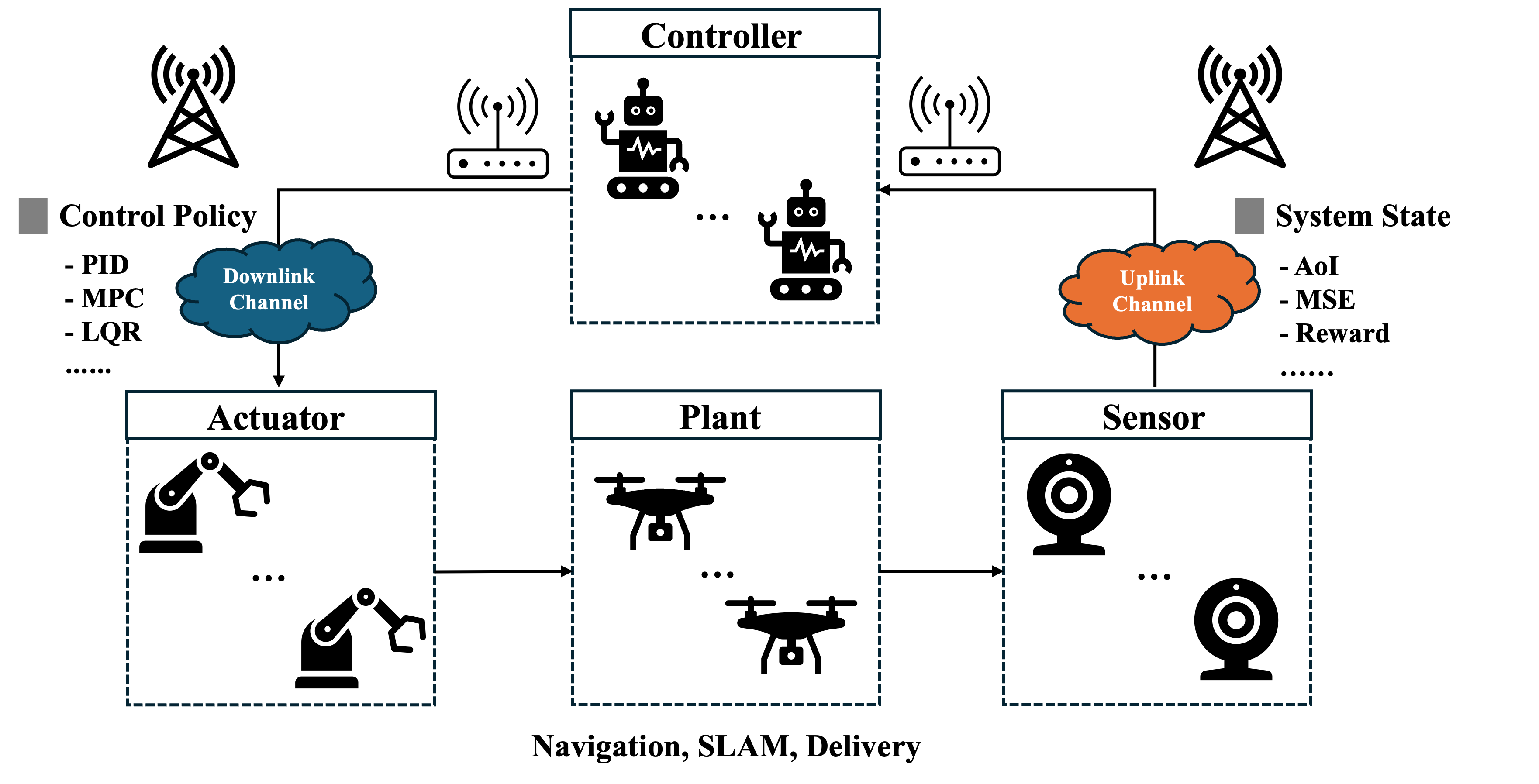}
    \caption{The framework of WNCS.}
    \label{WNCS}
\end{figure}

As shown in Fig. \ref{WNCS}, a typical WNCS is composed of plants, sensors, actuators, and a controller. The sensors transmit the measured data through the uplink channels to the controller, which processes the data, generates the control signal, and sends it to the actuators through the downlink channels. The design of WNCS is inherently multifaceted, integrating aspects of signal processing for plant state estimation, control theory for optimal plant regulation, and communication theory for the reliable transmission of sensor data. 

Control and networking systems interactively affect the performance of physical systems. For instance, in control systems, increasing the sampling rate is a common strategy to maintain system stability during physical disturbances. However, this approach can generate substantial network traffic, potentially impairing network performance and consequently degrading control system performance. From a communication perspective, network systems often employ conservative modulation schemes, characterized by fewer bits per symbol, to enhance packet delivery rates in sub-optimal communication conditions. Despite the goal of improving reliability, these conservative modulation schemes result in higher data volumes that must be transmitted. This additional burden on the network can reduce packet delivery rates, ultimately compromising control system performance even further. Therefore, coordinated development of estimation, control, and communication algorithms is crucial for superior control performance and communication under resource constraints, ensuring mutual optimization and integration.

Although the concept of control-communication co-design in WNCS was proposed decades ago \cite{park2017wireless}, most research has adhered to separate design principles, treating the controller and communication network independently. Traditionally, the communication domain has focused on enhancing performance metrics such as latency and reliability. Conversely, the control systems domain has focused primarily on control and estimation algorithm design, assuming predetermined communication policies. As WNCS become increasingly intelligent and tightly coupled, these separate design approaches are no longer viable. The integration of data-driven methods in the co-design of WNCS can facilitate the development of adaptive and resilient control strategies that handle the complexities and uncertainties inherent in modern control systems. By leveraging large datasets and sophisticated learning algorithms, data-driven approaches enable the continuous improvement of control and communication protocols, ensuring optimal performance even under varying conditions and constraints.

For instance, many existing studies assume perfect sensor measurements or communication links, which is often unrealistic in dynamic or poorly understood systems. Additionally, as the state and action spaces expand, traditional solutions like the MDP become impractical due to the curse of dimensionality. The application of data-driven techniques, particularly DL-based algorithms, is proving crucial in this co-design problem. These methods offer significant advantages by operating in a model-free manner, which is particularly beneficial in scenarios where precise knowledge of plant models is unavailable or infeasible to obtain.

In conclusion, achieving optimal system performance under resource constraints requires a holistic approach that integrates estimation, control, and communication algorithms, with a strong emphasis on data-driven methods. This integrated approach ensures mutual optimization and harmonious interaction, leading to more robust and efficient WNCS designs.

\begin{table*}[t]
    \caption{ML approaches for co-design.}
    \centering
    \label{tab: ML_co}
    \scriptsize
    \renewcommand{\arraystretch}{1.5}
    \resizebox{\textwidth}{!}{
        \begin{tabular}{|m{0.1\textwidth}<{\centering}|m{0.2\textwidth}<{\centering}|m{0.2\textwidth}<{\centering} |m{0.2\textwidth}<{\centering}| m{0.2\textwidth}<{\centering}|m{0.2\textwidth}<{\centering}|m{0.2\textwidth}<{\centering}|}
            \hline
            & Girgis et al.~\cite{girgis2020predictive}  &  Ji et al.~\cite{ji2023intelligent} &  Wu et al.~\cite{wu2020q} & Li et al.~\cite{li2019learning} & Girgis et al.~\cite{girgis2021predictive} &  Ji et al.~\cite{ji2023intelligent1} \\ \hline
            Scenario & Wireless sensor network control & Wireless sensor network control & Resource management & Resource management & Dynamic scheduling & Dynamic scheduling \\ \hline
            ML Method & Gaussian process regression & Bayesian learning & Parameter learning & Sequential learning & Gaussian process regression & Graph  learning \\ \hline
            Pros (+) & Supports more systems without compromising control stability & Achieves better overall performance in personalized production & Developed algorithms improve convergence and are easier to implement & Significantly reduces access latency and improves spectrum utilization & Achieves lower average control error and maintains control stability & Achieves up to 27.9\% performance improvement and 38\% runtime reduction \\  \hline   
            Cons (-) & Scalability could be further improved & Needs further exploration of distributed sensing and control co-design & Needs extension to general channels and multiple sensors & Rely heavily on accurate prediction and historical data & Requires frequent early-phase scheduling and accurate GPR prediction & Requires further exploration of multi-layer distributed sensing and control co-design\\ \hline
        \end{tabular}
    }
\end{table*}

\subsection{ML for Co-design}

In the burgeoning field of control communication co-design, ML techniques have emerged as a pivotal approach to address the intricate challenges of resource management, channel estimation, and dynamic scheduling. These methods enabled intelligent systems to adapt and optimize their performance in the face of complex and dynamic wireless environments, which are summarized in Tab.~\ref{tab: ML_co}.


The article by Girgis et al.~\cite{girgis2020predictive} addressed the challenges of remote control over wireless connections for scalable control systems. They introduced a novel scheduling algorithm to reduce the AoI for the last received states, ensuring timely observations. For unscheduled sensor-actuator pairs, a predictive control algorithm using Gaussian process regression (GPR) was proposed, leveraging historical data. The credibility of GPR predictions was linked to AoI, prompting a co-design approach integrating predictive control with AoI-based scheduling within the Lyapunov optimization framework. Simulations showed the method's superiority in achieving lower control errors compared to round-robin scheduling. Additionally, they addressed wireless resource allocation challenges by introducing a GPR-enhanced predictive control algorithm~\cite{girgis2021predictive}. This paper explored joint uplink-downlink scheduling and power allocation for multiple systems over wireless links, proposing a solution using GPR for state and action prediction to efficiently utilize resources. The GPR's credibility was determined by AoI, measuring data freshness. The paper formulated a network-wide problem to minimize the average AoI and transmission power, subject to communication reliability and control stability. A novel dynamic control algorithm using the Lyapunov optimization framework was proposed, demonstrating the ability to control more actuators stably compared to existing methods.

Zhang et al.~\cite{zhang2020stochastic} addressed security challenges in WNCSs by proposing a stochastic game framework. This framework tackled security concerns in WNCSs, focusing on denial-of-service (DoS) attacks. The authors constructed a two-player zero-sum stochastic game to model the interaction between a sensor and an intelligent DoS attacker. An improved LQG cost function incorporating energy efficiency was proposed. The paper discussed equilibrium strategies for continuous and discrete power settings, offering an algorithm to solve for these strategies. An example illustrated the approach's effectiveness in enhancing WNCSs' security and performance. Ji et al.~\cite{ji2023intelligent} introduced a framework for co-designing edge sensing and control in industrial cyber-physical systems, focusing on unknown models. They proposed a learning-based approach that analyzed model learning error to bound actual control performance, linking this bound to sensing design with relaxed assumptions about initial states and unknown system orders. The Cloud-Edge Symphony algorithm addressed the co-design problem, considering the limitations of single-edge computing units. Sensing, control, and learning processes were integrated for global optimization, and the algorithm was applied to personalized production in laminar cooling systems, demonstrating improved performance over previous methods.


Wu et al.~\cite{wu2019learning} explored optimal scheduling for remote state estimation under uncertain channel conditions, formulating scheduling problems with soft or hard communication rate constraints. Using dynamic programming and assuming known channel statistics, they presented structural results on the optimal policy. The authors proved the Q-factor's monotonicity and submodularity, leading to threshold-like structures in both problems. They developed stochastic approximation and parameter learning frameworks to address scheduling with unknown channel statistics, designing specialized learning algorithms and proving convergence. Performance improvements over standard Q-learning were demonstrated through numerical examples, which also discussed a method based on recursive estimation of channel quality. Li et al.~\cite{li2019learning} proposed a learning-based pre-allocation scheme called DPre to address low-latency access in industrial wireless networks. The scheme explored the correlation of devices' access behavior and utility diversity through sequential learning, enabling flexible pre-allocation decisions in time and frequency domains. A temporal-spatial utility metric ensured that more informative devices were prioritized for resource allocation. Theoretical analysis and simulations validated high spectrum utilization and effective packet preallocation, improving latency-critical services in industrial settings.

Together, ML-based approaches are revolutionizing control communication co-design, tackling challenges like resource management and channel estimation through predictive algorithms, such as GPR, and intelligent frameworks. Notable works demonstrated the efficacy of these methods in optimizing system performance, from achieving efficient wireless sensor network operations to enhancing security and latency in various networked control systems.

\subsection{RL for Co-design}
\begin{table*}[t]
\centering
\caption{RL approaches for co-design (COM-TO-CTRL).}
\label{table_rl_codesign_1}
\scriptsize
\begin{tabularx}{0.95\textwidth} { 
  | >{\centering\arraybackslash}m{0.14\textwidth} 
  | >{\centering\arraybackslash}m{0.2378\textwidth}
  | >{\centering\arraybackslash}m{0.2378\textwidth}
  | >{\centering\arraybackslash}m{0.2378\textwidth}| }

 \hline
& Baumann et al. \cite{baumann2018deep} & Liu et al. \cite{liu2021deep1} & Leong et al.\cite{leong2023stability} \\
\hline
Co-Design Problem
 & Simultaneous learning of control and communication.
 & A self-adaptive DRL approach.
 & Channel selection for remote state estimation.
\\
\hline
RL Method & DDPG & DQN & Multi-armed bandits
 \\
\hline
State 
 & The AoI of the sensor's and the controller's packet of plant i.
 & Cluster ID, data source nodes, connectivity matrix and data arriving rate.
 & Not explicitly stated; implicitly involves packet reception probabilities and state estimation.
 \\ 
\hline
Action 
 & Communication decision-making.
 & Policy for adjusting the routing connectivity of each cluster head node.
 & Selection of communication channels.
 \\
\hline
Reward 
 & Communication performance.
 & The product of data throughput, arrival rate, and connectivity rate.
 & Minimizing estimation error covariance (implicit in minimizing estimation regret).
 \\
\hline
Pros(+) or Cons(-) 
& (+) No need for an analytical dynamics model;
(+) Signiﬁcant communication savings;
(-) Worse performance than that with an accurate model.
 & (+) Significant improvements in data flow performance compared to static benchmarks.
 & (+) Guaranteed stability of the estimation process;
(-) Requires balancing exploration and exploitation properly.
\\
\hline
\end{tabularx}
\end{table*} 

RL is increasingly adopted in control-communication co-design systems, particularly for WNCS, due to its ability to navigate the complex, dynamic environments in which these systems operate. By leveraging RL, co-design systems can dynamically optimize trade-offs between critical factors such as latency and accuracy, efficiently manage limited resources, and adapt to changing conditions without the need for explicit environment models. This approach not only enhances system performance and efficiency but also ensures flexibility and scalability, making it an indispensable tool in the evolving landscape of WNCS.

In the domain of RL-based control-communication co-design problems, they can be categorized into three primary areas: (1) enhancing control performance through communication policy improvement (COM-TO-CTRL); (2) improving communication performance through control policy enhancement (CTRL-TO-COM); and (3) the joint design of control-communication systems (Joint Design).

In the first scenario, the focus is on optimizing communication strategies to reduce latency, packet loss, and bandwidth usage, thereby improving the effectiveness and efficiency of control decisions. 

Baumann et al.\cite{baumann2018deep} presented a DRL-based approach to ETC\cite{heemels2012introduction}, aiming to optimize control performance with fewer signal updates in networked systems constrained by limited communication resources. Unlike traditional ETC strategies that depend on mathematical models and predefined designs, this research employed a DRL framework to autonomously learn control and communication policies. Demonstrating DRL's first application to ETC, the paper revealed its capability to derive optimal strategies in an end-to-end fashion, particularly benefiting nonlinear systems where conventional methods falter. The effectiveness of this approach was proven across various control tasks, showcasing its potential to outperform or match model-based ETC techniques, even in scenarios where traditional methods were hindered by model limitations or complexity. Liu et al.\cite{liu2021deep1} proposed a self-adaptive DRL approach, leveraging the flexibility of Wireless mesh network (WMN) to improve communication performance dynamically. The approach integrated a DQN with a self-adaptive clustering model, which adjusted network topology during training to optimize communication routes. The results show significant improvements in data flow performance compared to static benchmarks. 
Leong et al. \cite{leong2023stability} focused on remote state estimation of Gauss–Markov processes over unknown communication channels. They used multi-armed bandit techniques for channel selection that ensured system stability and sample efficiency. The authors introduced a stability-aware Bayesian sampling algorithm and provided performance bounds on estimation regret, demonstrating that their algorithms maintained stability and achieved logarithmic regret over time.

In the second scenario, the main focus is on developing control policies that minimize interference and maximize the reliability and quality of communication signals, ensuring robust and efficient data transmission.

\begin{table*}[t]
\centering
\caption{RL approaches for co-design (CTRL-TO-COM).}
\label{table_rl_codesign_2}
\scriptsize
\begin{tabularx}{0.94\textwidth} { 
  | >{\centering\arraybackslash}m{0.1\textwidth} 
  | >{\centering\arraybackslash}m{0.18\textwidth}
  | >{\centering\arraybackslash}m{0.18\textwidth}
  | >{\centering\arraybackslash}m{0.18\textwidth}
  | >{\centering\arraybackslash}m{0.18\textwidth}|}
 \hline
 & Liu et al. \cite{liu2018energy}\cite{liu2019distributed} & Ballotta et al. \cite{ballotta2022reinforcement}\cite{ballotta2023compute}  & Huang et al.\cite{huang2022learning} & Du et al. \cite{du2022multi} \\
\hline
Co-Design Problem
 & Achieve certain communication coverage performance.
 & Latency-accuracy trade-off.
 & Optimize DoS attack strategies in CPS.
 & Dynamic resource management in 6G in-X subnetworks.
\\
\hline
RL Method & DDPG & Q-Learning & DDQN & Soft AC.
 \\
\hline
State & Current coverage score, state, energy consumption.
 & Error covariance matrix.
 & The holding time variables.
 & RSSI at each channel, remaining payload, previous action, and remaining time budget.
 \\ 
\hline
Action & The flying direction and flying distance for each Unmanned aerial vehicle(UAV).
 & Policy for the i-th sensor to transmit raw or processed data.
 & Attack power allocation.
 & Selection of channel and power level for transmission.
 \\
\hline
Reward & The combination of coverage score, fairness, and energy consumption.
 & The combination of coverage score, fairness, and energy consumption.
 & Average trace of the error covariance matrix.
 & Maximize the chance of successfully accomplishing payload transmission.
 \\
\hline
Pros(+) or Cons(-) & (+) Maximizes a novel energy efficiency function while ensuring effective and fair communication coverage, and network connectivity.
& (+) Surpass common designs and static policies.
& (+) Can address the uncertainty problem;
(+) Higher data efficiency;
(+) Better versatility, scalability, and attack performance.
 & (+) Does not require global knowledge, reducing signaling overhead and computational burden;
(-) Complexity increases with the number of subnetworks, though mitigated by the proposed attention mechanisms.
\\
\hline
\end{tabularx} 
\end{table*}

Liu et al.\cite{liu2018energy} comprehensively considered coverage, fairness, and energy consumption, proposing a DDPG-based energy-efficient control to enhance both the coverage and performance in a communication network connected by a group of drones. They further proposed a fully-distributed control solution \cite{liu2019distributed}, which is more practical and energy-efficient. The proposed algorithm outperforms the Random and Greedy policies.
Ballotta et al. \cite{ballotta2022reinforcement}\cite{ballotta2023compute} considered a network of smart sensors for edge computing applications. Sensors can either send raw, inaccurate measurements to the base or refine them locally before transmission, which incurs additional processing delay due to constrained computation. The authors proposed a DQN method that dynamically allocates the policy to determine whether each sensor should send raw data or processed data. This method balances the trade-off between communication and computation latency to maximize overall network monitoring performance, as validated through numerical experiments in applications like the Internet of Drones and self-driving vehicles.

Huang et al.\cite{huang2022learning} investigated optimizing DoS attack strategies in CPSs using the RL method. In this system, sensors observe different dynamic processes and transmit data to a remote estimator through wireless channels, which are targeted by a DoS attacker aiming to reduce transmission rates and degrade estimation accuracy. The study formulated the attack optimization problems as MDPs and addressed them using a DDQN method. Unlike previous works assuming the attacker has complete system knowledge, this study considered an attacker with limited information, making the problem more realistic. The proposed DDQN-based algorithm reduces the action space to improve learning efficiency and incorporates auxiliary tasks inspired by model-based RL to enhance data efficiency and learning performance. Experimental results demonstrate the superiority of the proposed algorithms in different system settings compared to traditional methods. 
Du \textit{et al.}~\cite{du2022multi} proposed a MARL method for dynamic resource management in 6G in-X subnetworks, addressing the challenges posed by dynamic mobility and interference. The novel MARL architecture, GA-Net, integrated a hard attention layer and a graph attention network to model inter-subnetwork relationships and manage resources effectively. The framework used only the RSSI, eliminating the need for instantaneous channel gain information, and was shown to outperform traditional and MARL-based methods.


For the last scenario, we implement a joint design approach that simultaneously optimizes both control-communication policies, leveraging synergies to enhance the overall performance and robustness of the system. This approach is more challenging than the aforementioned approaches. 

Zhao et al.\cite{zhao2023deep} proposed a novel DL-based framework for the WNCS, considering both of the sensor's AoI states and dynamic channel states. The authors addressed the challenges of joint design for an estimator–controller–scheduler training algorithm by designing a DRL-based algorithm that incorporates both model-free data and model-based data for optimizing controller and scheduler performance. Additionally, an importance sampling algorithm enhances learning efficiency by considering data accuracy. Experiments demonstrate that the proposed approach significantly outperforms separate design methods and benchmark policies, offering substantial improvements in various WNCS scenarios. 

Lima et al. \cite{lima2022model} addressed the challenge of controlling multiple systems, such as industrial robots, over a shared wireless network prone to packet loss and latency, which can potentially caused instability or task failure. To mitigate these issues, the authors proposed a control-adaptive network performance framework that dynamically adjusts target reliability and latency based on control system state information. They formulated the problem as a constrained MDP and utilize DRL to train a target reliability policy, optimizing network resource efficiency while meeting control-specific constraints. The effectiveness of this approach was demonstrated through simulations of a robotic conveyor belt task, where the learned policies significantly improved resource allocation efficiency compared to traditional control-agnostic methods. The study shows the potential of DRL in achieving scalable and efficient wireless control systems for industrial applications. 
In~\cite{eisen2022communication}, the paper addressed the challenge of controlling industrial systems over a shared wireless channel using edge computing. The authors proposed a communication-control co-design paradigm that adapted the network's QoS and control actions to the dynamic needs of each plant. They introduced a modular learning framework that leveraged DRL to optimize co-design policies in a model-free manner, demonstrating its effectiveness through a robotic conveyor belt task. The approach is inherently scalable and network-agnostic, making it suitable for various modern wireless protocols.


 The authors of~\cite{tehrani2021federated} tackled the distributed control of Next Generation wireless networks, which were expected to support demanding applications like augmented reality and autonomous vehicles. They proposed a federated DRL approach where BS collaboratively trained a DNN by sharing model weights instead of data. The federated DRL was evaluated in two versions, value-based and policy-based, showing superior performance over distributed and centralized DRL in terms of network sum rate, communication overhead, and privacy preservation.  

\begin{table*}[t]
\centering
\caption{RL approaches for co-design (joint design).}
\label{table_rl_codesign_3}
\scriptsize
\begin{tabularx}{0.95\textwidth} { 
  | >{\centering\arraybackslash}m{0.14\textwidth} 
  | >{\centering\arraybackslash}m{0.2378\textwidth}
  | >{\centering\arraybackslash}m{0.2378\textwidth}
  | >{\centering\arraybackslash}m{0.2378\textwidth}| }
 \hline
 & Zhao et al. \cite{zhao2023deep} & Lima et al. \cite{lima2022model} & Eisen et al. \cite{eisen2022communication} \\
\hline
Co-Design Problem
 & Joint design for an estimator–controller–scheduler training algorithm.
 & Controlling multiple systems over a shared wireless network.
 & Dynamic adaptation of QoS and control inputs in wireless edge industrial systems.
\\
\hline
RL Method & DQN & Primal-Dual Deep RL & Soft AC.
 \\
\hline
State & Current plant state, channel states, AoI.
 & Noisy plant state, channel conditions.
 & State variable of each plant, observation signal, QoS parameters.
 \\ 
\hline
Action & Control signals sent to the actuators or scheduling decisions for transmissions.
 & Control signals sent to the actuators and a resource allocation decision.
 & Selection of control inputs and QoS parameters.
 \\
\hline
Reward & The sensor transmission energy and Control performance.
 & Control performance and energy consumption.
 & Minimize the control cost and network resource cost.
 \\
\hline
Pros(+) or Cons(-) & (+) Co-design frameworks for WNCS with AoI and channel states;
(+) Develop novel schemes for enhancing the stability of training.
(+) Data accuracy is accounted for enhancing learning efﬁciency.
& (+) The optimality loss resulting from the use of policy parametrizations can be bounded.
 & (+) Can be applied across a wide range of modern wireless protocols and edge deployments;
(-) Practically challenging for modern control systems such as robotics due to non-linear dynamics and qualitative task performance metrics.
\\
\hline
\end{tabularx}
\end{table*}

The application of RL methods in control-communication co-design problems is experiencing rapid growth. A detailed comparison among the mentioned methods is shown in Tables \ref{table_rl_codesign_1} to \ref{table_rl_codesign_3}. However, the majority of current research focuses on transforming classical co-design problems into MDP frameworks, which are then addressed using RL techniques. Despite this progress, most works are still in the early stages. Challenges remain in scaling these approaches to complex, real-world systems and ensuring robust performance across diverse scenarios. Further research is needed to address these issues and fully realize the potential of RL in enhancing control-communication co-design.

\subsection{Conclusion for Co-design}
In this section, we have explored the significant advancements in the field of learning-based control and communication co-design. ML techniques have emerged as a crucial approach to tackle the intricate challenges of resource management, channel estimation, and dynamic scheduling in complex wireless environments. Noteworthy works have demonstrated the effectiveness of predictive algorithms, such as GPR, and intelligent frameworks in optimizing system performance, enhancing security, and reducing latency in various networked control systems. RL has also gained prominence in the co-design of control-communication systems, particularly in WNCS. RL's ability to navigate complex, dynamic environments has enabled co-designed systems to dynamically optimize trade-offs between latency and accuracy, efficiently manage limited resources, and adapt to changing conditions without explicit environment models. DRL has been applied to ETC, demonstrating its potential to derive optimal strategies in an end-to-end fashion, especially benefiting nonlinear systems where conventional methods struggle.

\subsection{Open Issues and Future Work}
In the future, the integration of ML and RL in control communication co-design is expected to continue driving innovation and performance improvements. As the complexity of wireless environments and the demand for efficient resource management grow, these learning-based approaches will be instrumental in developing intelligent, adaptable, and scalable systems. Future research may focus on further refining predictive algorithms, exploring the potential of FL for distributed control, and developing more efficient and robust RL techniques. Moreover, the application of learning-based control communication co-design is poised to expand across various domains, from industrial cyber-physical systems and the IoT to autonomous vehicles and beyond. As 6G networks emerge, the need for dynamic resource management and interference mitigation will become even more critical, and multi-agent RL methods are expected to play a vital role in addressing these challenges. Research directions include combining scheduling schemes with event-triggered strategies for efficient resource utilization, incorporating transmission power control to mitigate packet dropouts, fadings, and failures, and exploring protocol sequences to develop analytically simpler and practical communication scheduling protocols. Additionally, investigating computation-control co-design algorithms for low delay and high accuracy in time-critical applications, as well as communication-control co-design for connected vehicle systems facing dynamic channel conditions and frequent handoffs, will be crucial. Finally, methods for reconfiguring controllers and communication topology to mitigate cyber-attacks and maintain system performance will enhance security and resilience. By addressing these future research directions, the field of learning-based control and communication co-design will continue to evolve, meeting the growing demands and challenges of next-generation wireless networks and diverse application domains.